\documentclass[10pt,twocolumn,letterpaper]{article}
\pdfoutput=1 

\usepackage[pagenumbers]{cvpr} 

\usepackage[]{graphicx}
\usepackage{amsmath}
\usepackage{amssymb}
\usepackage{booktabs}
\usepackage{tabularray}    
\usepackage[x11names]{xcolor}
\usepackage[export]{adjustbox}
\usepackage{tikz}
\usepackage{soul}

\usepackage[pagebackref,breaklinks,colorlinks]{hyperref}
\usepackage{xcolor}
\usepackage{xspace}

\newcommand{\ProjectAria}{Project Aria\xspace}
\newcommand{\AriaDevice}{Project Aria device\xspace}
\newcommand{\AriaDevices}{Project Aria devices\xspace}

\usepackage[capitalize]{cleveref}
\crefname{section}{Sec.}{Secs.}
\Crefname{section}{Section}{Sections}
\Crefname{table}{Table}{Tables}
\crefname{table}{Tab.}{Tabs.}

\title{Project Aria: A New Tool for Egocentric Multi-Modal AI Research}
\author{{Jakob Engel, Kiran Somasundaram, Michael Goesele, Albert Sun, Alexander Gamino, Andrew Turner, Arjang Talattof, Arnie Yuan, Bilal Souti, Brighid Meredith, Cheng Peng, Chris Sweeney, Cole Wilson, Dan Barnes, Daniel DeTone, David Caruso, Derek Valleroy , Dinesh Ginjupalli, Duncan Frost, Edward~Miller, Elias Mueggler, Evgeniy Oleinik, Fan Zhang, Guruprasad Somasundaram, Gustavo Solaira, Harry~Lanaras, Henry Howard-Jenkins, Huixuan Tang, Hyo Jin Kim, Jaime Rivera, Ji Luo, Jing Dong, Julian Straub, Kevin Bailey, Kevin Eckenhoff, Lingni Ma, Luis Pesqueira, Mark Schwesinger, Maurizio Monge, Nan Yang, Nick Charron, Nikhil Raina, Omkar Parkhi, Peter Borschowa, Pierre Moulon, Prince Gupta, Raul Mur-Artal, Robbie Pennington, Sachin Kulkarni, Sagar Miglani, Santosh Gondi, Saransh Solanki, Sean Diener, Shangyi Cheng, Simon Green, Steve Saarinen, Suvam Patra, Tassos Mourikis, Thomas Whelan, Tripti Singh, Vasileios Balntas, Vijay Baiyya, Wilson Dreewes, Xiaqing Pan, Yang Lou, Yipu Zhao, Yusuf Mansour, Yuyang Zou, Zhaoyang Lv, Zijian Wang, Mingfei Yan, Carl Ren, Renzo De Nardi, Richard Newcombe}\\[2mm]
Meta Reality Labs Research}
\date{June 2023}

\begin{document}

\maketitle

\begin{abstract}
   Egocentric, multi-modal data as available on future augmented reality (AR) devices provides unique challenges and opportunities for machine perception. These future devices will need to be all-day wearable in a socially acceptable form-factor to support always available, context-aware and personalized AI applications. Our team at Meta Reality Labs Research built the \AriaDevice: An egocentric, multi-modal data recording and streaming device with the goal to foster and accelerate research in this area. In this paper, we describe the \AriaDevice{} hardware including its sensor configuration, the corresponding software tools, and the available machine perception functionalities that make it the ideal tool for egocentric machine perception and contextual AI research.
\end{abstract}

\section{Introduction}
\label{sec:intro}

All-day wearable AR glasses promise to be the next big paradigm shift in computing: They can lift interaction with the digital world from 2D screens into the 3D world around us; blending seamlessly into our lives as opposed to diverting our attention into small 2D rectangles held in our hands. Yet, in order to be more than mere 3D versions of 2D screens, they also require a new compute- and interaction-paradigm that is context-aware, highly personalized and natural to interact with. Creating this new paradigm is a significant challenge that still requires a broad range of research to be solved.
Fortunately, the recent breakthrough of internet-trained Large Language Models such as GPT4~\cite{openai2023gpt4} and llama2 \cite{touvron2023llama} promise to solve a major part of this challenge, making it a lot more tractable:
They demonstrate that modern Transformer architectures combined with sufficient training data enable both long-range reasoning and information retrieval while also adopting seamlessly to new tasks using a local context window and prompt engineering. Taken together, this enables human-level interaction with digital agents that blend into how we humans interact naturally. 

\begin{figure}
    \centering
    \includegraphics[width=\linewidth]{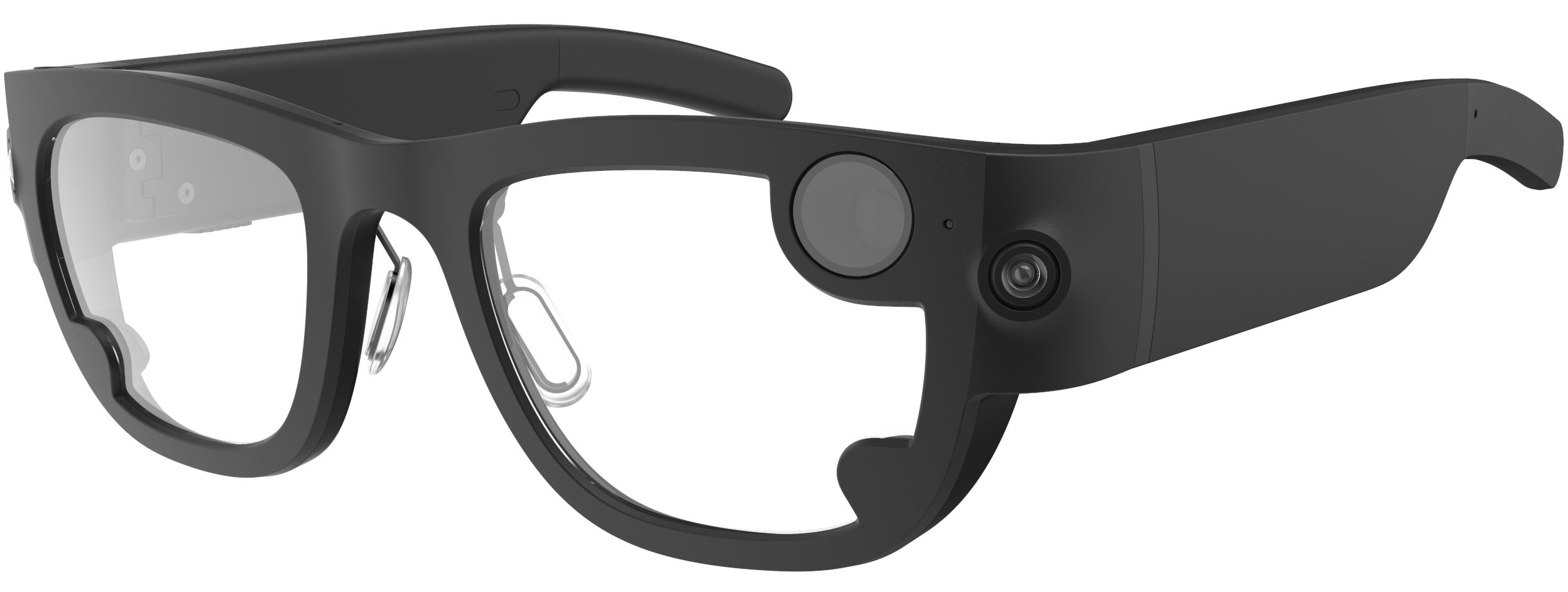}
    \caption{The \AriaDevice.}
    \label{fig:aria-image}
\end{figure}
\subsection{\ProjectAria}

However, we believe that this paradigm shift in personalized computing requires another crucial ingredient: the ability for AI Agents to adapt to the unique, personal context and preferences of the \textit{individual user}. Some of this can be achieved by leveraging a persons digital footprint -- the aggregate of interactions with the internet through phones and laptops. It is only a matter of time for such personalized AI Assistants to emerge. Yet, our digital footprint only represents a small fraction of the experiences that matter to us, and rarely contain the most important ones, which play out in the real world, are highly personal and often take place in unpredictable places and at unpredictable times. 

Importantly, this applies not only on an individual level, but also in aggregate: The wealth of digital data accumulated over 25 years in the digital realm only represents a small -- and often severely biased -- fraction of the sum total of human experience. For example, the overwhelming majority of images available on the web and used to train AI models were consciously captured with handheld devices and curated before upload. Any outtakes, any data that is not deemed interesting is typically deleted or edited although it arguably represents the majority of situations we encounter in our daily life.

A direct consequence of this is that many state of the art methods in the space of Machine Perception and AI (DALL·E2\cite{ramesh2022hierarchical}, GPT-4~\cite{openai2023gpt4}, DreamFusion~\cite{poole2022dreamfusion}, SAM~\cite{kirillov2023segment}, MaskRCNN~\cite{he2017mask}, CLIP~\cite{radford2021learning}, DINOv2~\cite{oquab2023dinov2}) excel when applied to allocentric 2D images and viewpoints, but fare comparatively poorly at tasks that involve egocentric data or require structured reasoning and understanding in 3D/4D space. While  this is partially due to the increase in data/compute requirements, it is also a direct result of the aforementioned shortcomings of data typically  
available through web platforms.

In order to address this gap and to enable the next big paradigm shift towards context-aware, personalized, and human-oriented AI, we have created \textit{\ProjectAria}: At its core, the \AriaDevice{} is a data-capture system in glasses form factor that is sufficiently light and unobtrusive to be worn for long time-spans without inhibiting natural activities and behavior (see Figure~\ref{fig:aria-image}), allowing to capture \textit{ecologically valid} data. The device features a rich multi-modal sensor suite that approximates what can be expected in future AR glasses for the purpose of environment- and user-understanding. The onboard battery allows the device to record 1-2 hours of data (with the nominal recording profile). Much longer recordings are possible with an external power bank.

In this technical report, we introduce the \AriaDevice{} as well as the software and Machine Perception Services that come with it. We make these available to research institutions around the world to foster advancements in the field of egocentric perception towards personalized AI. 
We also summarize the principles and standards that we have established and adopted to protect the privacy of both wearers and bystanders in accordance with Meta's Responsible Innovation Principles \cite{metari}.

Since its launch in 2020, \AriaDevices{} have been used by research groups in the USA, UK, Switzerland, India, Canada, Singapore, Colombia and Japan. In 2022, we released the Aria Pilot Dataset~\cite{ariapilot}. 
To find out more about \ProjectAria{} and how to become a research partner, please visit our website~\cite{ariawebsite}.

This report is organized as follows: Section \ref{sec:device} introduces the \AriaDevice{} and its capabilities. Section \ref{sec:recordingtools} introduces the software and tools required for recording and using recordings for research. Section \ref{sec:mps} enumerates the set of first-level machine perception capabilities we offer through a web service in order to accelerate and simplify use of \ProjectAria{} data. Section \ref{sec:privacy} details the privacy and responsible innovation principles we have established, and how the \AriaDevice{} implements them. Section \ref{sec:applications} gives a set of example research applications that are enabled by Project Aria. We end this paper with a conclusion in Section \ref{sec:conclusion}.

\section{Device}
\label{sec:device}

\subsection{Sensor Suite}
\begin{figure*}[t]
    \includegraphics[width=\linewidth]{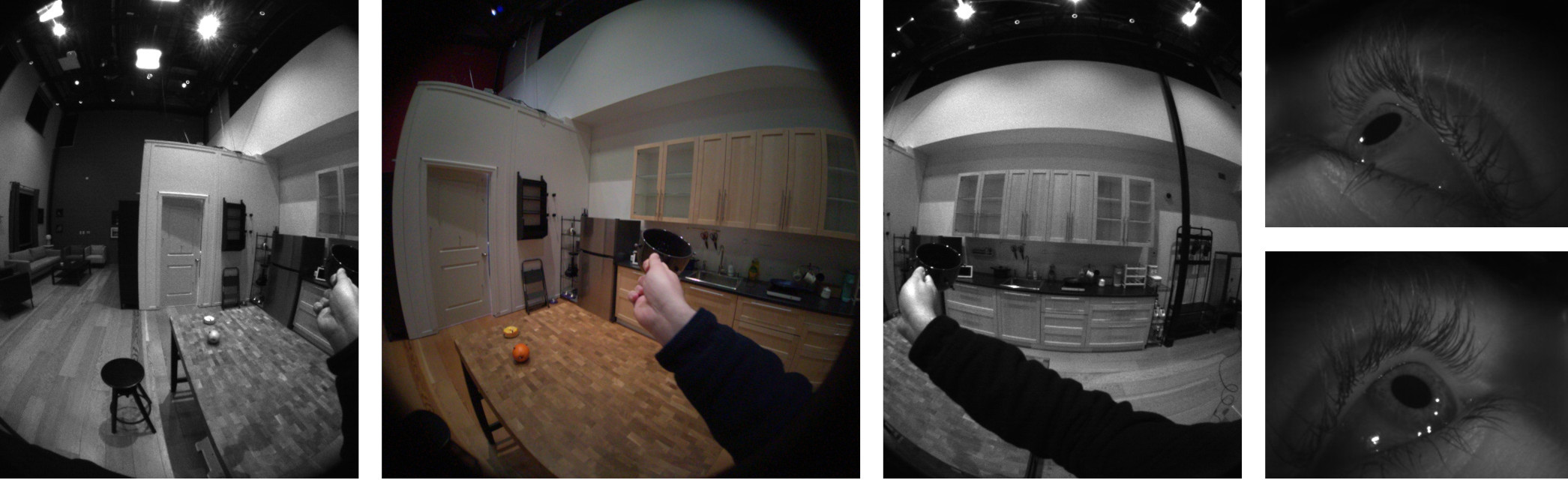}
    \caption{Example images from the \AriaDevice{} cameras. Left to right: left Mono Scene camera, POV (RGB) camera, right Mono Scene camera, two Eye-tracking cameras. Output of POV (RGB) and Mono Scene cameras are rotated for visualization.}
    \label{fig:aria-multi-modal-scene-cameras}
\end{figure*}

\begin{figure}[t]
    \centering
    \includegraphics[width=\linewidth]{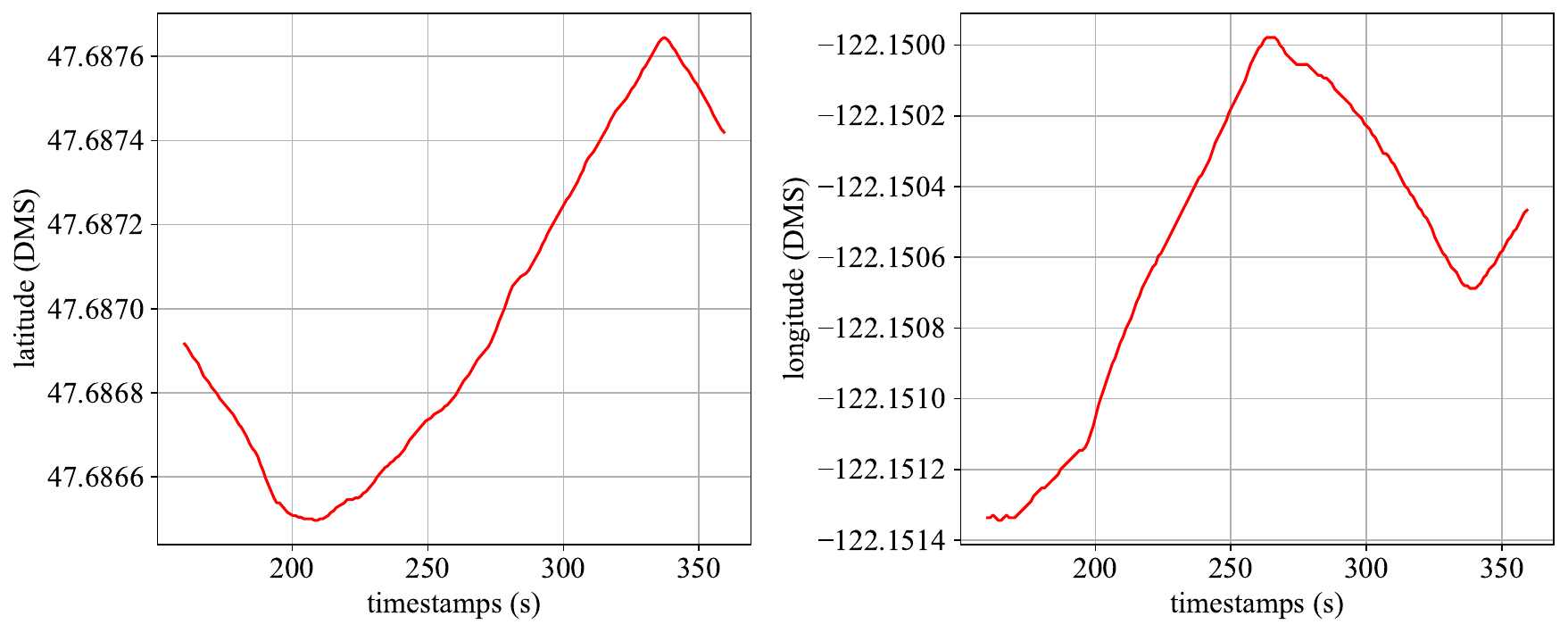}
    \includegraphics[width=0.49\linewidth]{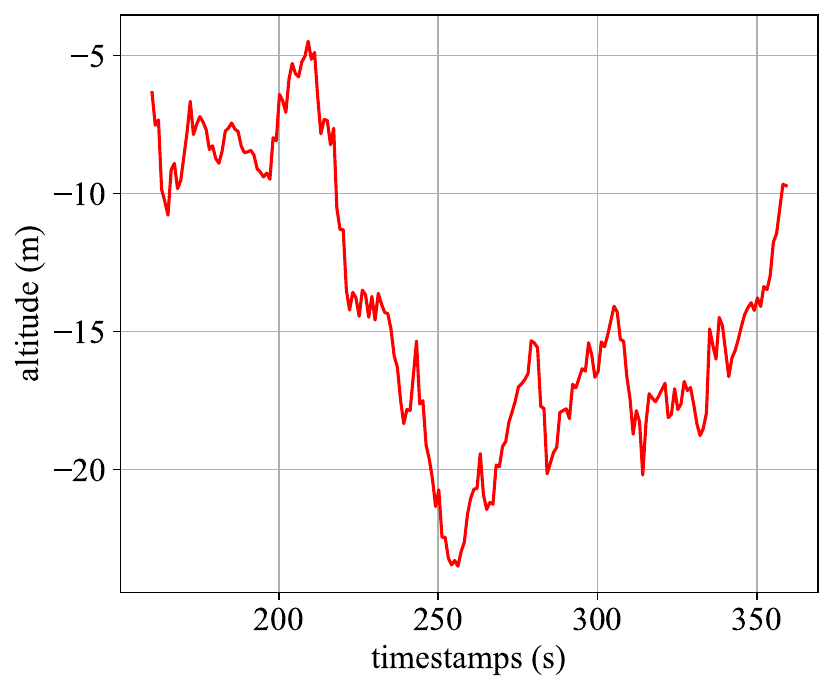}
    \includegraphics[width=0.49\linewidth]{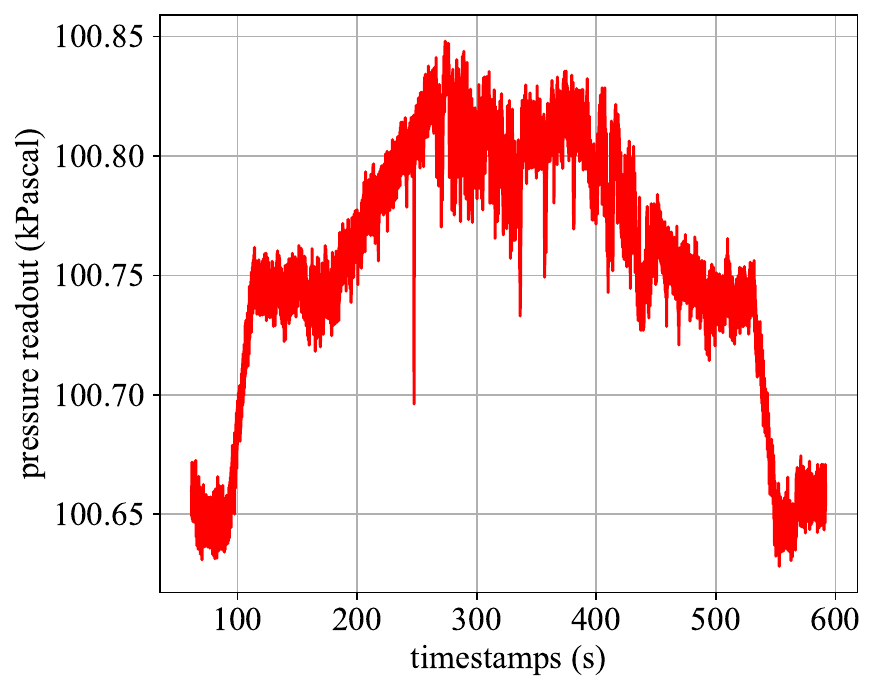}
    \includegraphics[width=0.49\linewidth]{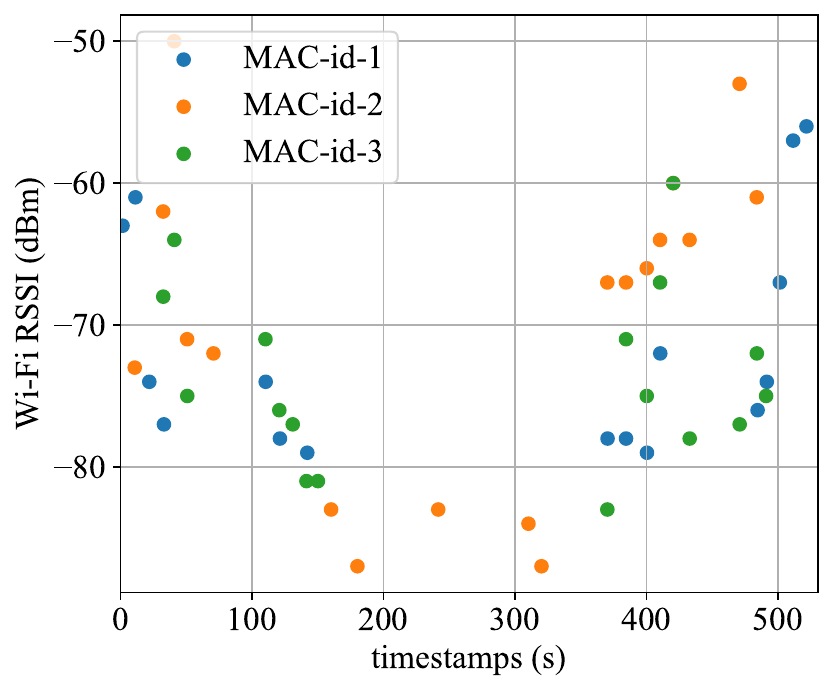}
    \includegraphics[width=0.49\linewidth]{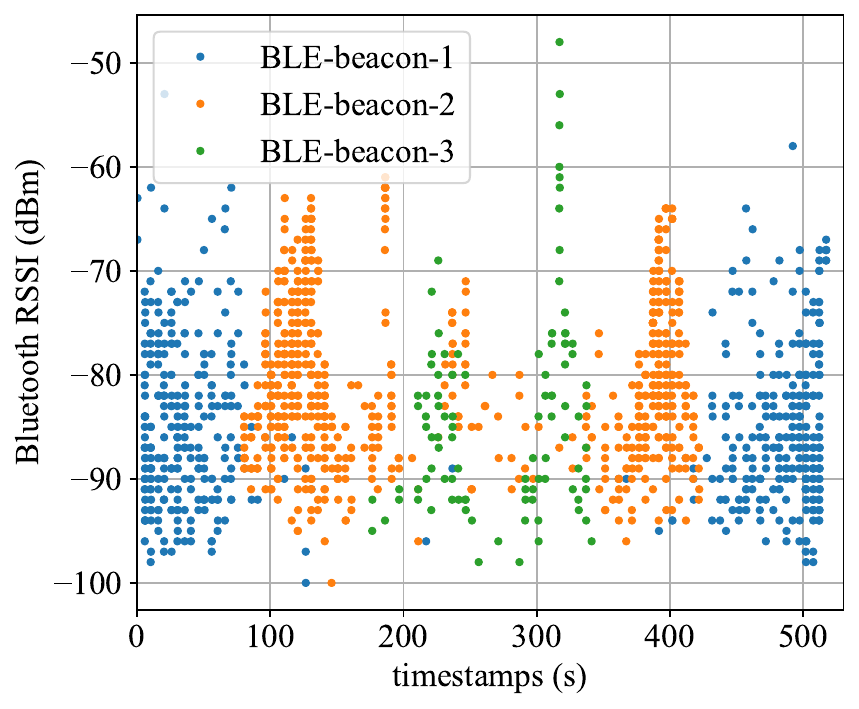}
    \caption{Illustration of the GNSS and Wi-Fi/Bluetooth sensor data. Top to bottom:  GNSS signal (individual plots of latitude, longitude, altitude), pressure measurements provided by the barometer,
    signal strengths from different sources as recorded by the Wi-Fi and Bluetooth receivers. 
    }
    \label{fig:aria-multi-modal-gps-wps}
\end{figure}

We built the \AriaDevice{} to emulate future wearable devices catering to machine perception rather than human consumption. To mimic the sensor stack needed for machine perception capabilities on these future devices, we integrate a rich suite of sensors that record egocentric multi-modal data (see 
Figures~\ref{fig:aria-multi-modal-scene-cameras}\,--\,\ref{fig:aria-multi-modal-gps-wps}).

Sensor streams are tightly calibrated and time-aligned to make some problems fundamentally easier to solve.  This is aligned to future expectations of wearable device hardware, but introduces new challenges - such as the lack of Optical Image Stabilization (OIS) or Auto-Focus (AF). 

To satisfy the strict requirements for capturing representative data in a wearable form-factor constraints, we built the \AriaDevice{} as a data-collection and streaming device. Specifically, it is not designed to handle on-device computation workloads.

\subsection{Form Factor and Fit}
The \AriaDevice{} is designed to contain a rich set of sensors while being light (around 75g) and socially acceptable. The devices provide a good fit for a broad range of the population with two device sizes. They come with adjustable nose pads and temples (i.e.~the temple tips can be bent inwards or outwards to improve fit). The glasses temple have a lot more flexibility than a conventional pair of glasses, which means the small size can stretch up to what many glasses might call a medium or large size.

The sensors on the \AriaDevice{} have been chosen to (1) approximate what we expect to be available on future all-day-wearable AR glasses, and (2) fit into the strict size, weight and power envelope required by the target form-factor to obtain ecologically valid data. We selected a broad range of sensors beyond just cameras as we believe multi-modal egocentric data to be key to solve various Machine Perception and AI tasks -- see Section~\ref{sec:applications} for some examples. The following gives an overview of what sensors are available on a \AriaDevice\ (see also Figure \ref{fig:aria-glasses-blowup}):
\begin{itemize}
  \setlength\itemsep{0em}
\item \textbf{Mono Scene Cameras}: Two monochrome, global-shutter cameras on the left and right side of the glasses. They have a horizontal field of view (HFOV) of 150\textdegree\ and are angled outwards to maximize peripheral vision while allowing for some stereo overlap. These cameras are used for supporting machine perception capabilities such as Visual SLAM~\cite{engel2014lsd, mur2017orb, mourikis2007multi}.
They have a resolution of 640x480 pixels and use Fisheye (F-Theta) lenses.
\item \textbf{Point of View (POV) RGB Camera}: A single high-resolution, rolling-shutter RGB camera on the left side of the glasses with a HFOV of 110\textdegree. The RGB camera also uses a F-Theta Fisheye lens, has a maximum resolution of 2880x2880 pixels and faces forward with an approximately 4\textdegree\ bias towards the ground. 
\item \textbf{Eye Tracking Cameras}: Two monochrome, global-shutter, inward-facing cameras for eye tracking with a diagonal field of view (DFOV) of 80\textdegree. They typically operate at a 320x240 pixels resolution.
\item \textbf{IMUs}: Two inertial measurement units (IMU), one on each side of the glasses. The left IMU is configured to sample at 800\,Hz with saturation limits of 4g (accelerometer) and  500\textdegree/s (gyroscope). The right IMU samples at 1000\,Hz with saturation limits of 8g and 1000\textdegree/s. We intentionally chose different IMU models so that their higher-order error behaviors are more likely to be uncorrelated.
\item \textbf{Microphones}: A microphone array comprised of 7~microphones distributed around the glasses (5 front, 1 on each side), allowing to capture spatial audio at 24\,bits with a configurable sample rate of up to 48\,kHz.
\item \textbf{Magnetometer}: A magnetometer located on the rim of the glasses to minimize electromagnetic interference, measuring the ambient magnetic field (3-axis) with a resolution of 0.1\,$\mu$T and a sample rate of 10\,Hz.
\item \textbf{Barometer \& Thermometer}: A barometer sensor capturing local air pressure and temperature at a resolution of 0.66\,Pa and 0.005\textdegree\,C, respectively, with a sample rate of 50\,Hz.
\item \textbf{GNSS receiver}: A global navigation satellite system receiver supporting GPS and Galileo constellations, providing pseudo-range measurements as well as lat/long/height solutions with a sample rate of 1\,Hz.
\item \textbf{Wi-Fi \& Bluetooth transceiver}: A Wi-Fi and Bluetooth radio with the ability to regularly scan and record received signal strengths (RSSI) from Wi-Fi beacon frames (both 2.4G and 5G) and from Bluetooth beacons. Scanning is nominally performed at 0.1\,Hz.
\end{itemize}

\begin{figure}
    \centering
    \includegraphics[width=\linewidth]{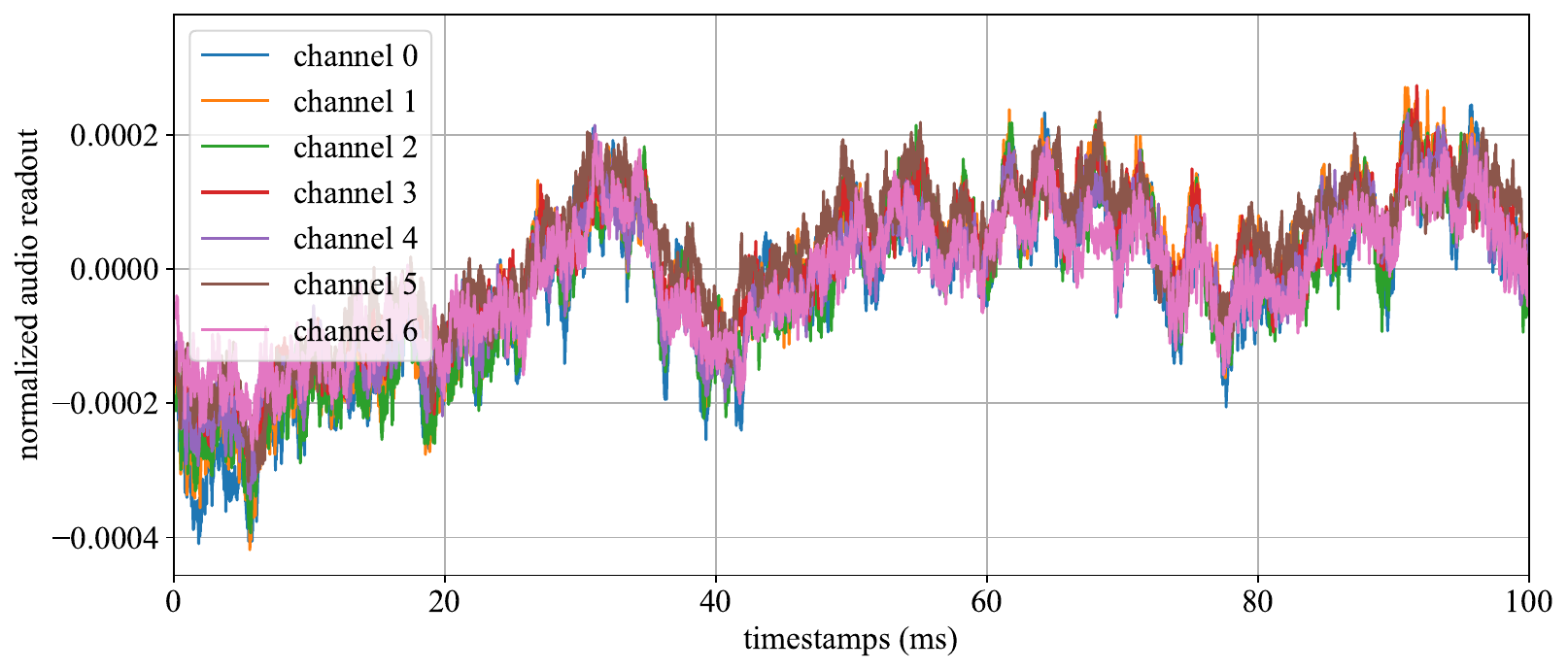}
    \caption{Example time-series data from the multi-channel microphone array on the \AriaDevice{}. Audio data is saved in 32-bit format and normalized in the range of $[-1, 1]$.}
    \label{fig:aria-multi-modal-audio}
\end{figure}

\begin{figure}[t]
    \centering
    \includegraphics[width=\linewidth]{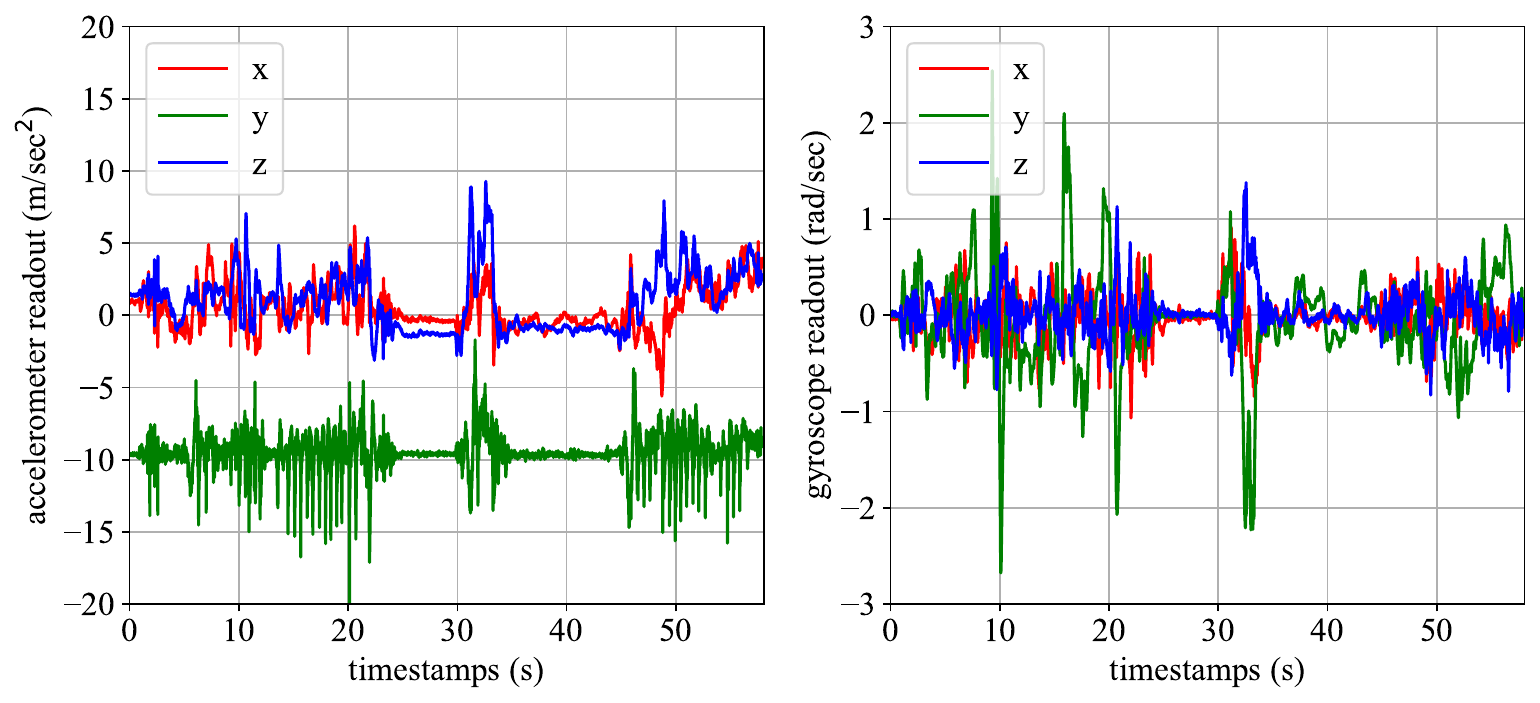}
    \includegraphics[width=\linewidth]{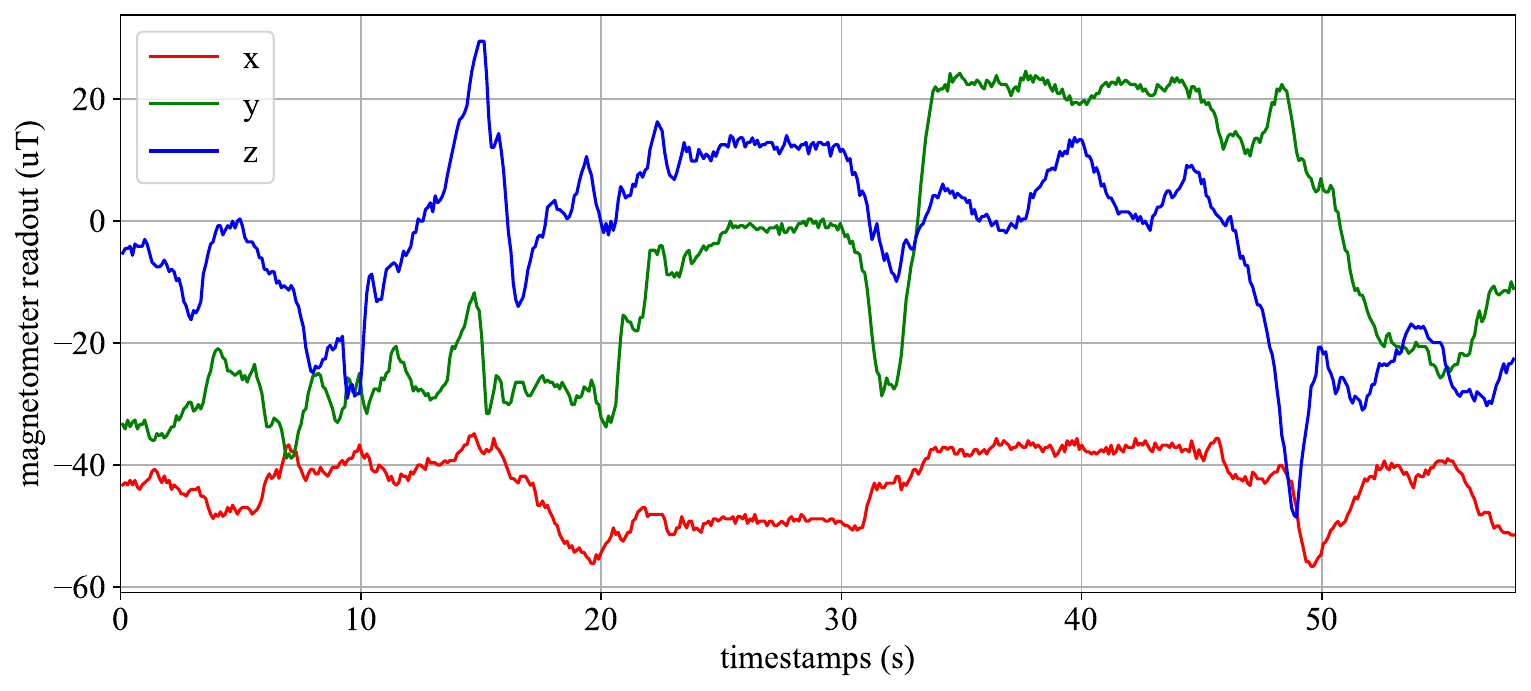}
    \caption{Example data from various motion sensor and location signal data. Top to bottom: accelerometer and gyroscope data provided by the IMUs, and magnetic field measurements provided by the magnetometer.}
    \label{fig:aria-multi-modal-data-imu-mag-baro}
\end{figure}

Many of these sensor settings can be configured through \textit{recording profiles} that are selected at the start of a recording and allow to modify camera frame-rate and resolution, as well as to enable or disable different sensor streams. This is important as recording all sensors at maximum resolution and rate is not feasible due to power and bandwidth limitations. Furthermore, disabling sensor streams or reducing their rate and/or resolution can be desirable from a privacy perspective, and is an effective strategy to prolong maximum recording time. Available recording profiles are listed in the  \ProjectAria{} documentation site \cite{ariadocs}.

\subsection{Mounting and Rigidity}
Precise 6DoF alignment -- that is, relative positioning and orientation between all sensors -- is important for many basic machine perception algorithms, and the \AriaDevice{} has been designed to facilitate this. All sensors are mounted onto a magnesium frame spanning the front of the glasses. With the most non-rigid portion being the nose-bridge, there are two primary sensor clusters on the left and right side of the glasses. Sensors within the same cluster have a strong rigid connection.

We provide extrinsic and intrinsic calibration parameters for each sensor computed at manufacturing time (factory calibration). Through our Machine Perception Services (MPS, see Section \ref{sec:mps}) we additionally make more accurate online-calibration parameters available that account for the small deformations/changes that might occur while wearing the glasses. Please refer to \cite{ariadocs} for more details.

\subsection{Time and Time-alignment}

The \AriaDevice\ is designed to allow accurate timestamping 
of all sensor data with respect to a local on-device time source. This is essential to combine different modalities in downstream machine perception tasks.

In addition, \AriaDevices{} provide the ability to align and convert local timestamps to a time domain shared across multiple devices. Accurately translating to a common time domain is critical when combining or comparing data from different sources and devices. 
In our sample datasets, we use SMPTE~LTC~timecode~\cite{smptet} to provide a common accurate time domain across multiple \AriaDevices{} (see Aria Pilot Dataset \cite{ariapilot}). For situations where lower accuracy is acceptable we have also implemented a methodology for sharing a common time domain over Wi-Fi leveraging the TicSync timing protocol~\cite{conf/icra/HarrisonN11}. For both methodologies, the inner working of the time sharing mechanism is handled at the device level. This means, from the perspective of a user of the data, every sensor reading simply includes a timestamp in the aligned time domain in addition to a timestamp in the local device time.
Please refer to the  \ProjectAria{} documentation \cite{ariadocs} for more details on the exact definition and conventions for time\-stamps of the different sensor modalities.

\begin{figure*}[pt]
    \centering
    \includegraphics[width=0.95\linewidth]{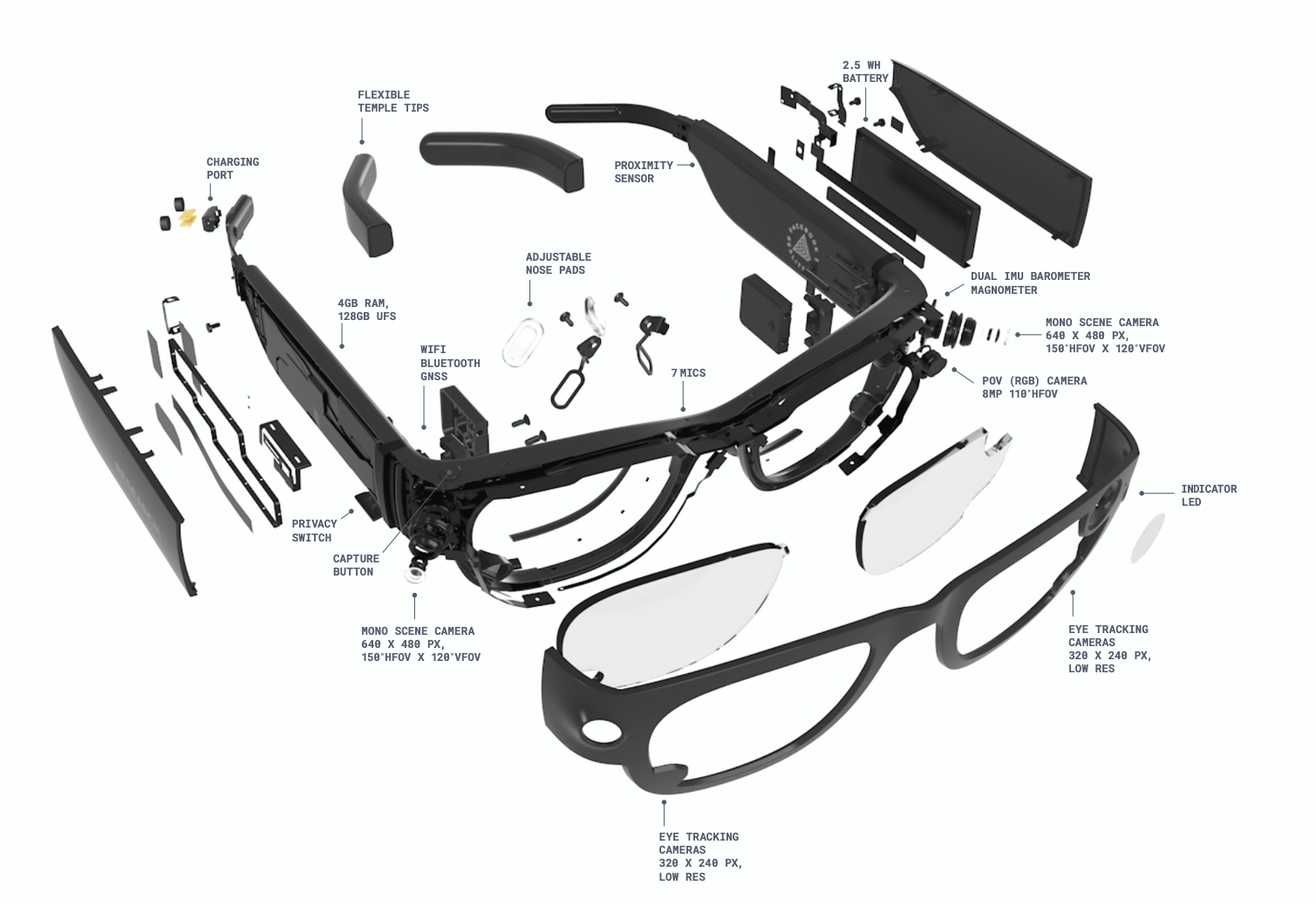}
    \caption{\AriaDevice{} hardware overview of the components, the various sensors, switches, LEDs, battery, etc.}
    \label{fig:aria-glasses-blowup}
    \vspace{5mm}
\centering
    \includegraphics[frame,height=0.4663560112\linewidth]{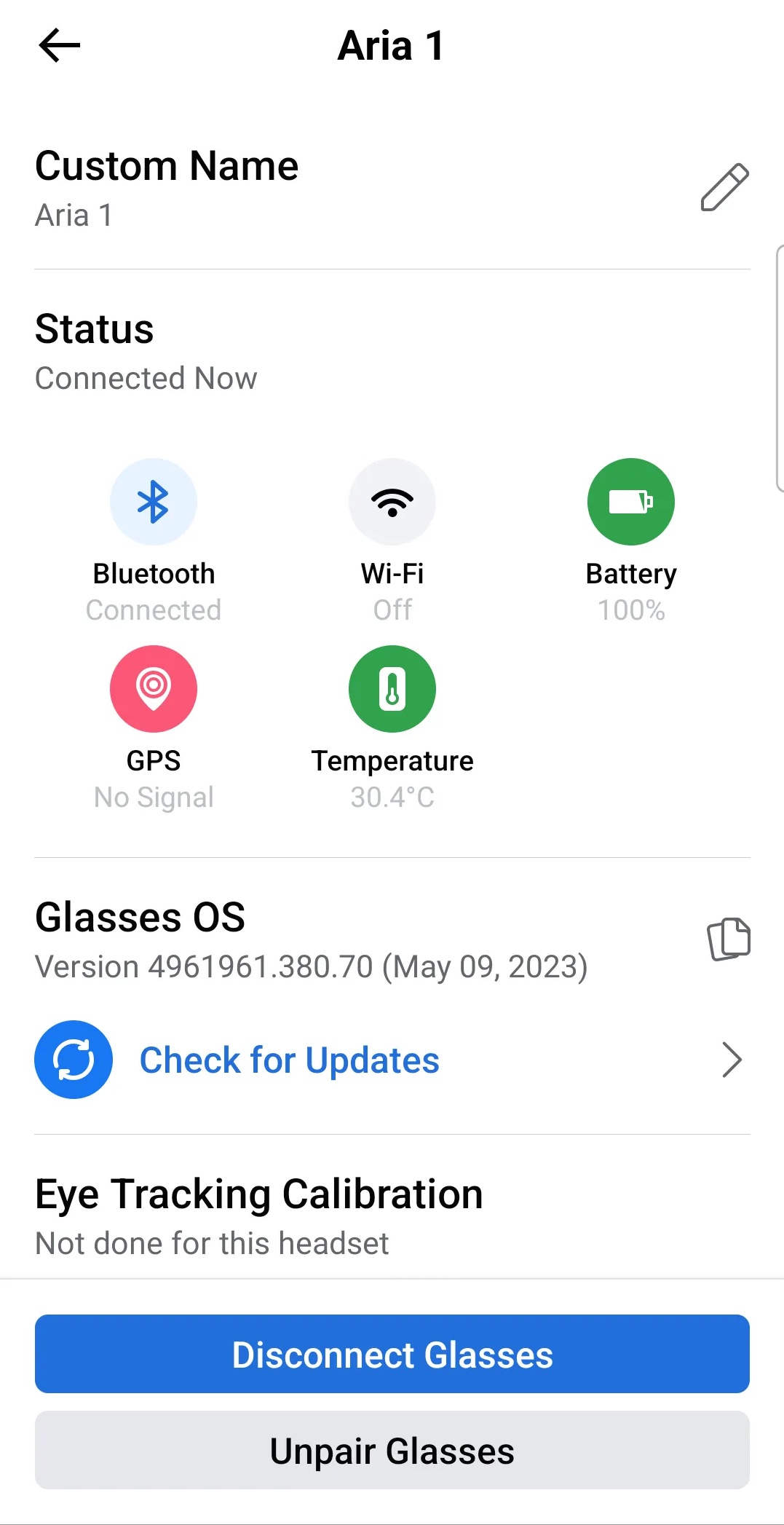}\hfill
    \includegraphics[frame,height=0.4663560112\linewidth]{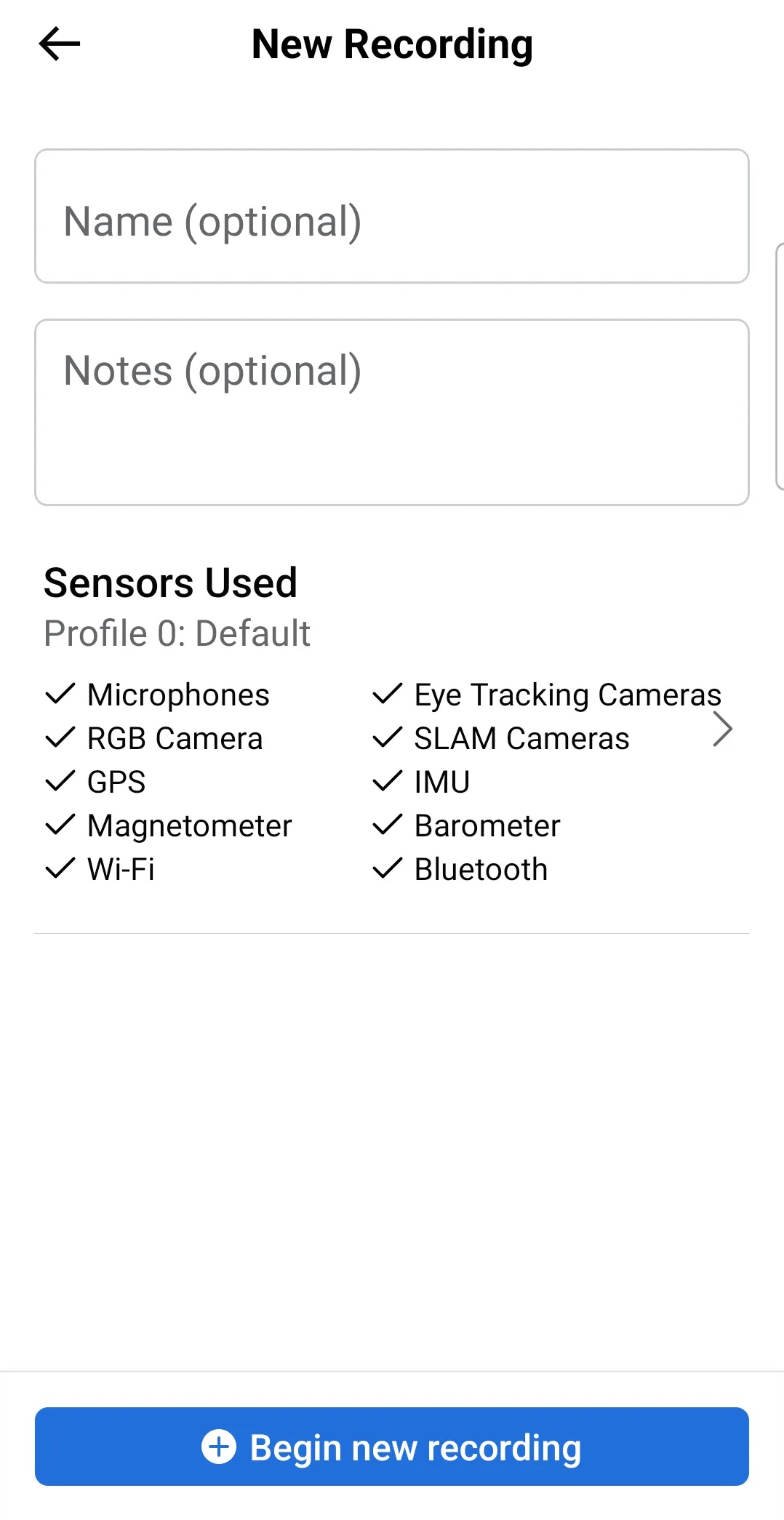}\hfill
    \includegraphics[frame,height=0.4663560112\linewidth]{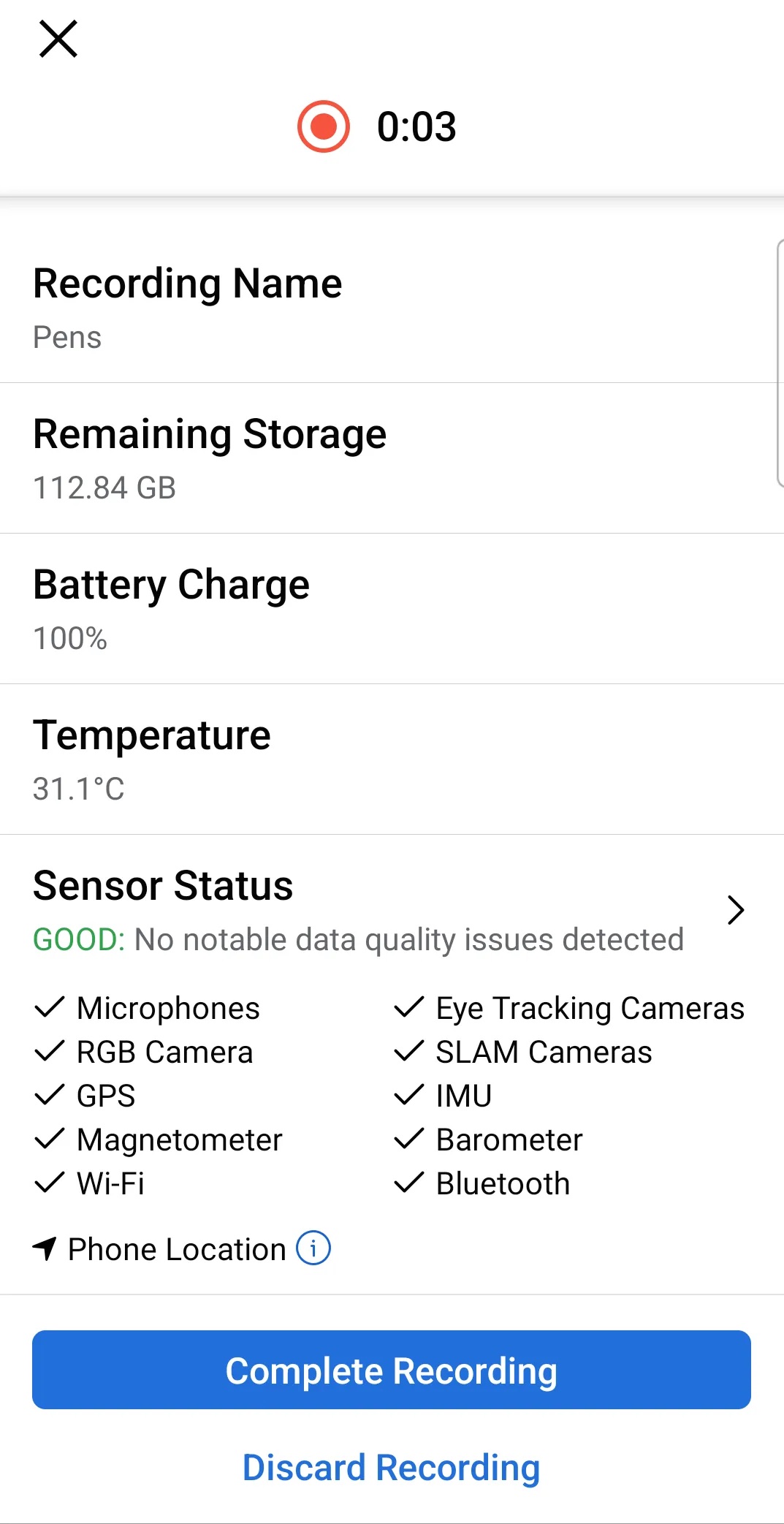}\hfill
    \includegraphics[frame,height=0.4663560112\linewidth]{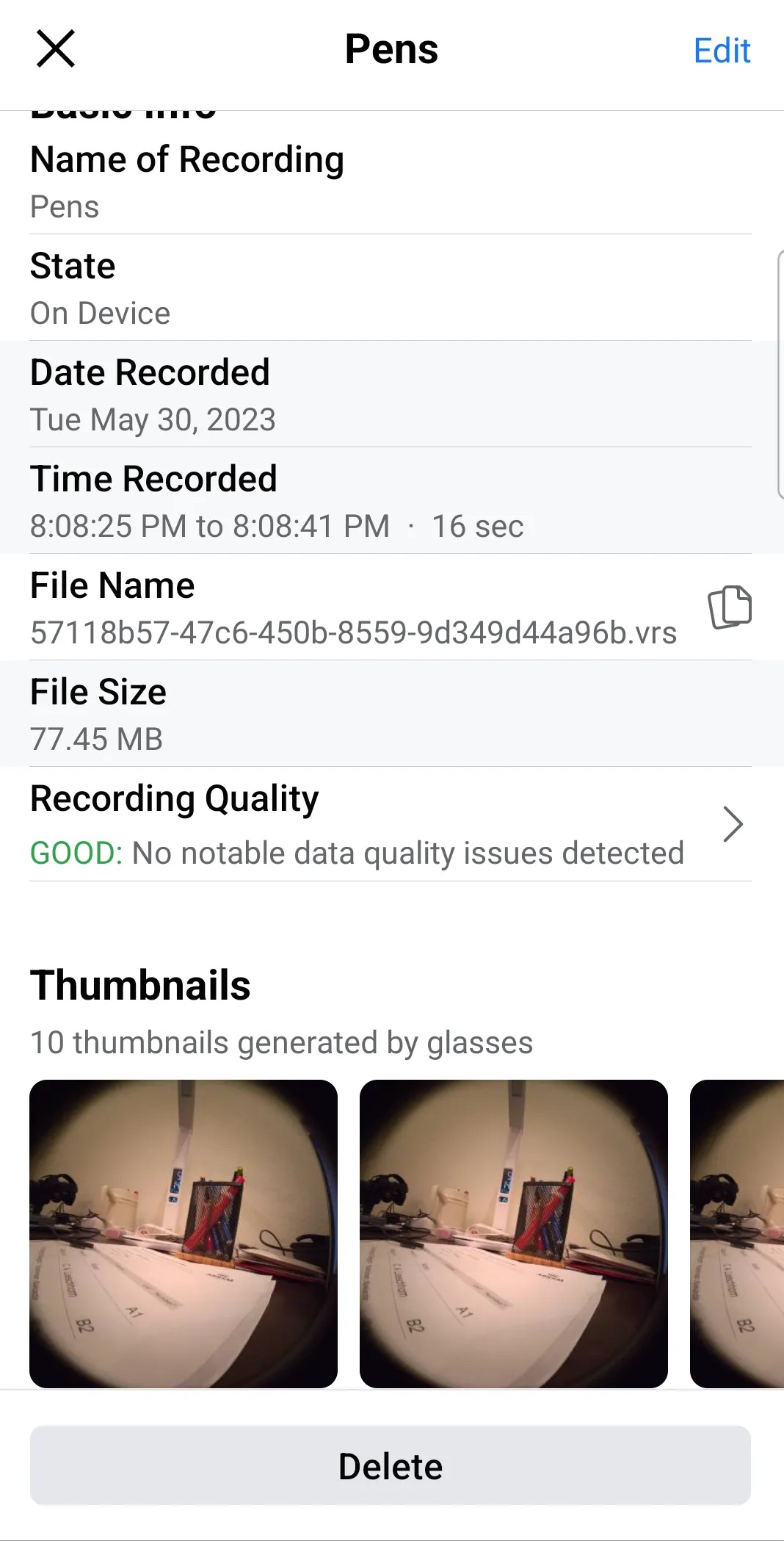}
    \caption{Basic functionalities of the mobile phone companion app. From left to right, the screenshots illustrate a) the device status information, b) the recording profile that configures the sensors before recording, c) status of the on-going recording, d) thumbnail preview of the recordings visible after a recording finishes}
    \label{fig:companion-app}
\end{figure*}

\section{Recording Tools}
\label{sec:recordingtools}

The primary interface to interact with \AriaDevices{} is a mobile phone companion app. Recordings can be initiated and stopped via the app or the device's capture button.
The app is also used to set the sensor configuration by selecting a recording profile. The profile controls which sensors record, at what frequency and resolution, as well as the output format (e.g., storing images in RAW or JPEG format). 
Multiple recording profiles are available to tailor for different research use cases. Additional profiles are being added as necessary. Once a recording is made on the device, thumbnail previews are available on the companion app for convenient review of the captured data. These functionalities are illustrated in Figure~\ref{fig:companion-app}.

Once the recording is complete, the user can download the recorded data from a \AriaDevice{} via an USB connection to a local machine for further processing. A user can optionally upload their recordings to our Machine Perception Services (MPS), which apply state-of-the-art processing to recover device trajectories, online calibration, a semi-dense point cloud and eye gaze information (see Section~\ref{sec:mps} for details).

All device sensors are recorded in VRS file format~\cite{vrsdocs}.  We selected VRS as the data container because it is an open file format designed to record and playback streams of AR sensor data and because it supports very large file sizes. The VRS files contain streams of time-sorted records generated for each sensor, with one set of sensors per stream.

In order to easily visualize and interact with data, we provide \ProjectAria{} tools as part of an open source repository \cite{ariatools}. This is a set of tools and libraries for accessing, visualizing and manipulating recordings from Aria. The C++/Python toolkit includes VRS Data Provider and Viewer interfaces, enabling researchers to read and visualize \ProjectAria{} sequences, to retrieve and interact with device calibration data, and to read and process the output of the MPS.
More details about these tools are available from documentation website \cite{ariadocs}.
The \ProjectAria{} tools are available to install from PyPI via
{\tt pip install projectaria\_tools} (see \cite{ariatools} for more details).

\section{Machine Perception Services}
\label{sec:mps}

We provide a range of foundational machine perception capabilities upon which research partners can build their projects. We expect these capabilities to be provided in a similar form on any future AR device.

These capabilities are exposed as Machine Perception Services (MPS) enabled by a set of proprietary algorithms that are designed for \AriaDevices{} and provide superior accuracy and robustness on the recorded data compared to current off-the-shelf open source solutions.

MPS is provided by post-processing VRS recordings on Meta's backend servers. To use the service, \ProjectAria research partners upload the recording, and, later on, can download the results. Furthermore, we include MPS output in public Aria-based datasets making it available to the broader community. 

Please refer to the \ProjectAria{} documentation site \cite{ariadocs} for more information about Aria MPS, the output format and their specifications, and an overview over the tooling available to visualize and make use of the data.

\begin{figure}[t]
    \centering
{\setlength{\fboxsep}{0pt}%
\setlength{\fboxrule}{0.5pt}%
\fbox{\includegraphics[width=0.99\linewidth]{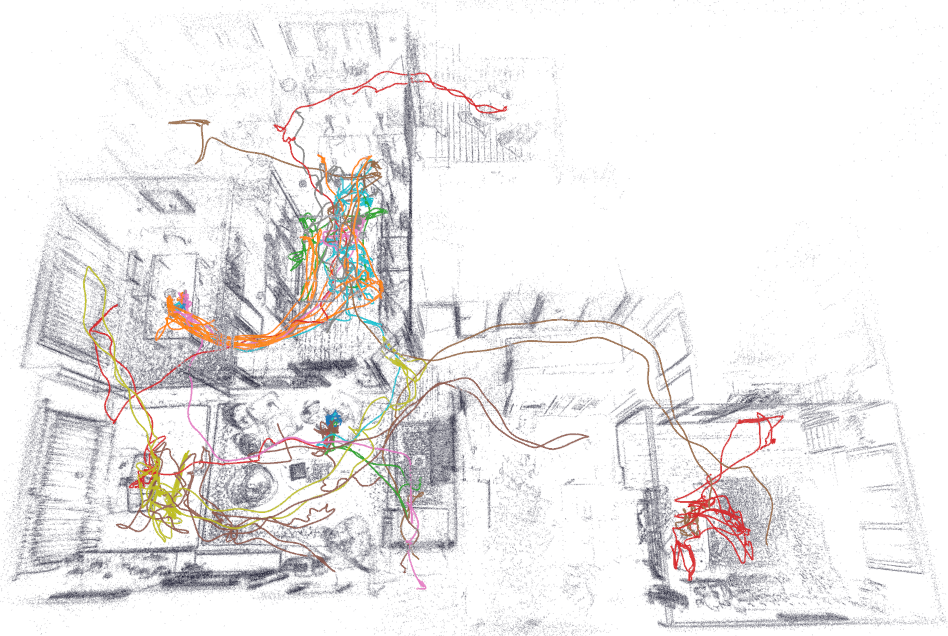}}}
    \caption{Closed-loop trajectories and semi-dense point clouds of 18 recordings of Aria Pilot Dataset \cite{ariapilot} collected in the same home space.}
    \label{fig:mps_static}
\end{figure}

\subsection{Trajectories}
\label{sec:trajectories}

A highly accurate 6-DoF device trajectory is the foundation to understand the geometric relation of the device and its wearer to the environment. Device trajectories are generated by a state-of-the-art VIO and SLAM system -- similar to what can be expected on future AR and VR HMD's -- followed by offline post-processing and refinement. We use multiple of the available sensors (including cameras, IMUs, GNSS, Wi-Fi, and barometer) to improve accuracy and robustness, and further take advantage of precise knowledge of the sensor models, timing, and rigidity of \AriaDevices{}. This allows us to robustly localize the device even under the often challenging conditions that occur with real-world data - such as fast motion, low or highly dynamic lighting, partial or temporary occlusion of the cameras, as well as a wide range of static and dynamic environments. 

We provide two types of trajectories as output, open loop and closed loop.  The \textit{open loop trajectory} is a high frequency (1\,kHz) odometry estimation, computed strictly causally with a real-time-compatible method. The accumulated translation drift of this open loop trajectory is no more than 0.4\% of the distance traveled, and usually significantly less. The \textit{closed loop trajectory} is a 1\,kHz trajectory estimated in post-processing. It is fully optimized and provides poses in a single frame of reference. We also provide the ability to jointly process multiple recordings, placing them into a common coordinate frame, as shown in Figure~\ref{fig:mps_static}. Closed loop trajectories have a typical global RMSE translation error of no more than 1.5\,cm in room-scale scenarios.

\subsection{Online Calibration}
\label{sec:onlinecalib}

\begin{figure}[tbp]
    \centering
    \includegraphics[width=\linewidth]{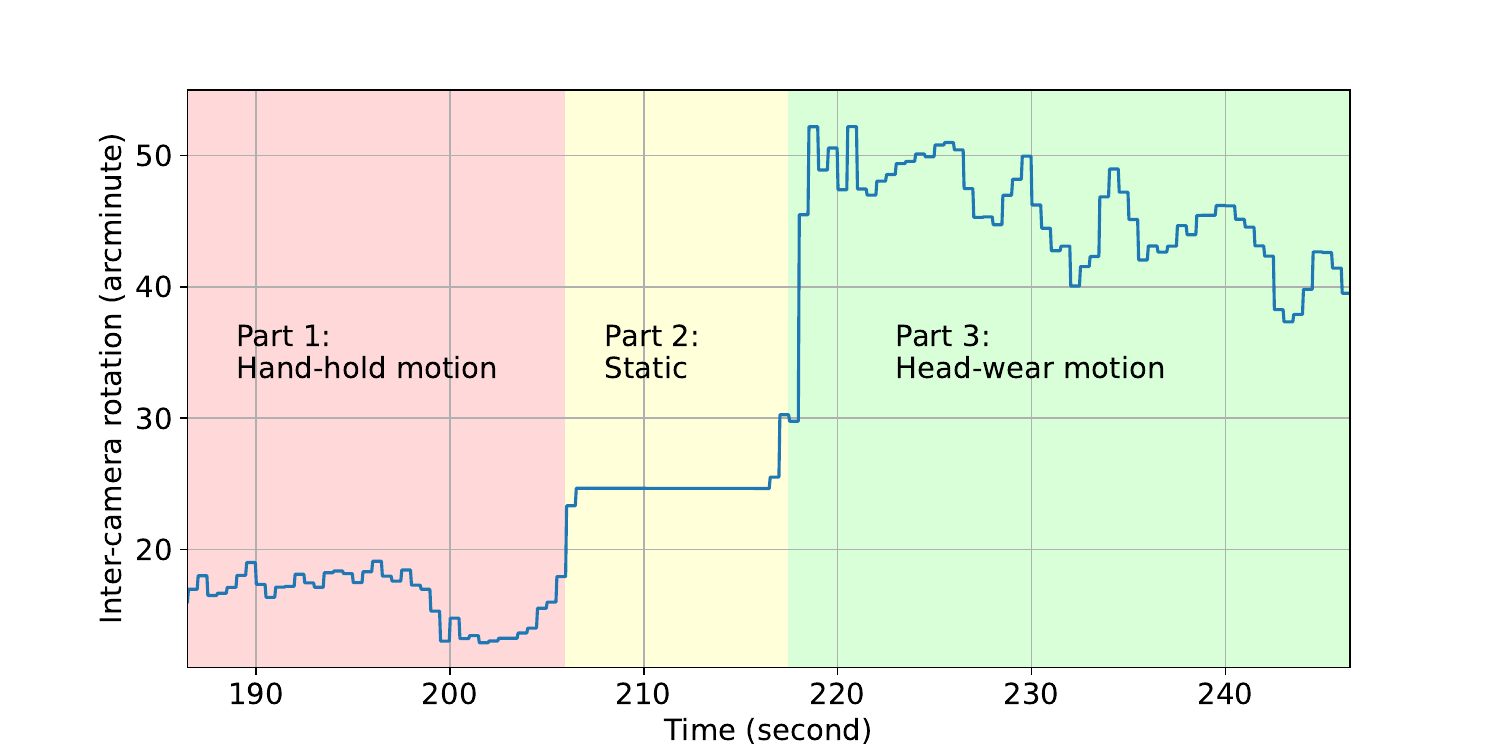}
    \caption{Example of rotational deformation between left and right Mono Scene cameras, estimated by MPS. The recording used for this figure contains 3 sections: hand-held motion, no motion, and head-worn motion. The effect of external force applied to the glasses when worn on the head is clearly visible.}
    \label{fig:mps_online_calib}
\end{figure}

Accurate device calibration is essential to enable high geometric accuracy for downstream 3D perception tasks.
Even though \AriaDevices{} are built to be as rigid as possible, device calibration parameters are not perfectly constant over time due to temperature changes, aging, and external forces applied to the device. 
Figure~\ref{fig:mps_online_calib} shows an example where taking off the device after wearing causes around 25\,arc\-min instantaneous rotational deformation between the left and right Mono Scene cameras (corresponding to roughly 1.5~pixel shift in the images). 
To account for this, we estimate the time-varying intrinsic and extrinsic calibrations of cameras and IMUs as part of MPS, and make the result available to researchers.

\subsection{Semi-Dense Point Cloud}
\label{sec:semi-dense}
To provide an intuitive understanding of the environment a recording was taken in, we compute semi-dense tracks and point clouds as part of MPS; see Figure~\ref{fig:mps_static} for an example.  Similar to the odometry trajectory, these tracks are computed causally and provide an accurate -- though partial -- reconstruction of the static portion of the environment. We also provide the sets of all 2D observations that were used to triangulate each 3D point. Tracks are obtained by continuously spawning new points in images-regions with high gradient, and tracking these over time and across the left/right Mono Scene camera using affine-invariant photoconsistency of local patches. Finally, the 3D point clouds are post-processed and placed into the global frame of reference that is defined by the closed loop trajectories.

\subsection{Eye Gaze Tracking}

Gaze direction is an important indicator of a wearer's attention, and will likely be one of the crucial inputs to context-aware, personalized AI agents.
We compute and provide eye gaze from the \AriaDevice{} eye tracking cameras, estimating a single per-frame 3D ray anchored to the central pupil frame, also called a cyclopean eye frame \footnote{The central pupil frame origin is defined at the midpoint between the left and right pupils.}. We also provide confidence intervals, as eye tracking accuracy can vary by situation and user. 

Furthermore, the Aria companion app described in Section \ref{sec:recordingtools} implements the option to capture personalized eye calibration - this allows to improve the accuracy of eye gaze tracking by compensating for user-specific biases. With the current model, we observe a median gaze ray error of 1.5\textdegree\ after applying the personalized calibration.

\begin{figure}[tbp]
    \centering
    \includegraphics[height=0.442\linewidth]{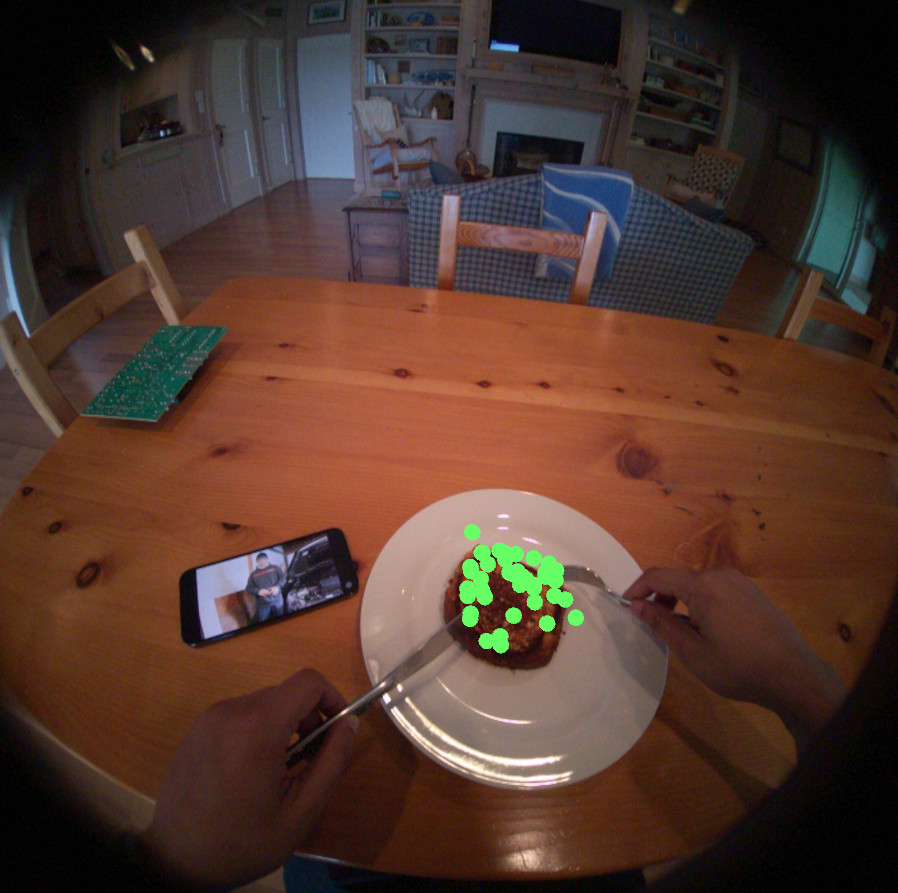}%
\hfill%
{\setlength{\fboxsep}{0pt}%
\setlength{\fboxrule}{0.5pt}%
\fbox{\includegraphics[height=0.442\linewidth]{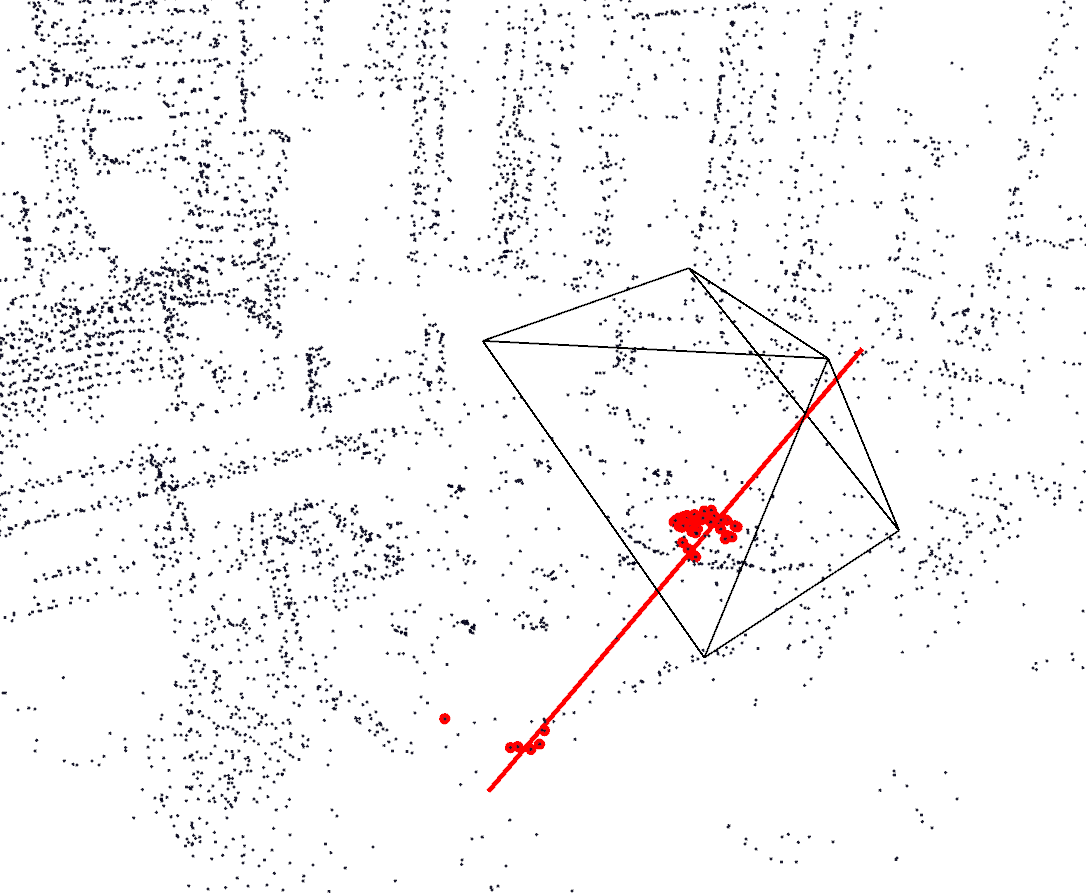}}}
    \caption{Eye gaze computed on a recording in which the user is looking at the food on their plate. Left: The RGB image with semi-dense points that are close to the gaze ray projected as green dots. Right: The \AriaDevice{} pose (shown as RGB camera frustum), semi-dense points, and eye gaze ray. Semi-dense points close to the gaze ray are highlighted in red.}
    \label{fig:mps_gaze}
\end{figure}

\section{Privacy Considerations}
\label{sec:privacy}

AR glasses and in general egocentric recording devices such as \AriaDevices{} promise to make technology more accessible, but also pose novel and unique challenges for security and privacy: the more they succeed in being unobtrusive, the more important it becomes to preserve and respect the privacy not only of the wearer, but also that of bystanders and individuals the wearer interacts with. 

A stated goal of  \ProjectAria{} is to pioneer responsible innovation for research leveraging egocentric data and devices, both by establishing guidelines and principles to preserve privacy of wearers and bystanders as well as by building privacy-facilitating features directly into the device where possible. 

Throughout the development of \ProjectAria{} we followed Meta’s Responsible Innovation Principles~\cite{metari}, which express our commitment to building inclusive, privacy-centric products. 

In concert with our principles we have designed and made available privacy-centric hardware and software features to research partners. The \AriaDevice{} has an LED indicator that signals to bystanders when the device is recording raw data. Furthermore, \AriaDevices{} have a privacy switch: When activated during a recording session, the device immediately stops and deletes the current recording. This allows a wearer to immediately and easily fulfill a request of a bystander to delete any audio or video recording that might have been captured of them. 

We require all \ProjectAria{} partners to follow the \ProjectAria{} community guidance \cite{communityguidelines} and to practice responsible research, protecting the privacy of those who wear the research devices, and most importantly, those who do not.

\section{Example Research Applications}
\label{sec:applications}
This section provides a brief overview of research tasks that leverage Project Aria's unique features, from low-level machine perception functions to high-level user- and environment-understanding. Project Aria is designed to enable and connect research across this spectrum: While the former benefits from well-calibrated and understood sensors and access to raw data, the latter can leverage the multiple modalities available or build upon the machine perception functions provided by Aria Machine Perception Services (MPS). 
Note that this is neither an exhaustive review of the respective fields nor a complete list of tasks that are enabled by Project Aria. It is meant to provide examples how Project Aria devices or Aria data can be leveraged, with an emphasis on Aria's unique combination of form-factor, sensors and Machine Perception Services.

\begin{figure}
    \centering
    
{%
\setlength{\fboxsep}{0pt}%
\setlength{\fboxrule}{0.5pt}%
\fbox{\includegraphics[width=0.325\linewidth]{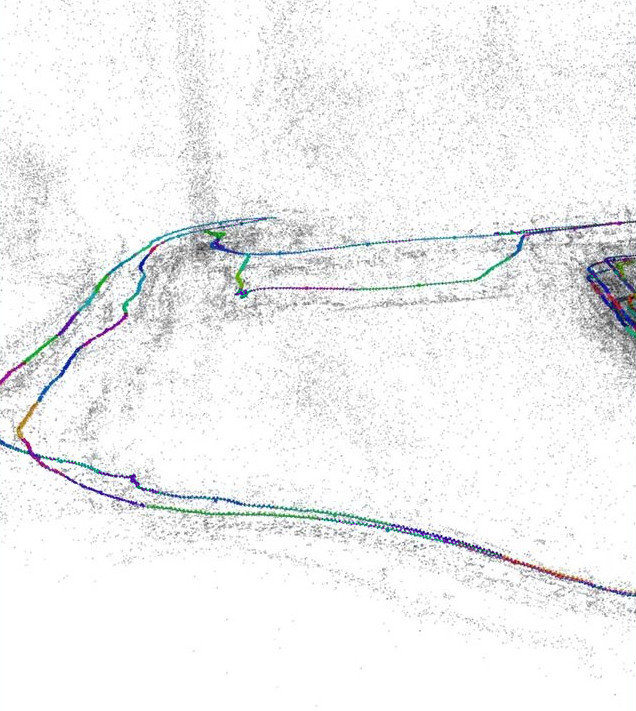}}\hfill%
\fbox{\includegraphics[width=0.325\linewidth]{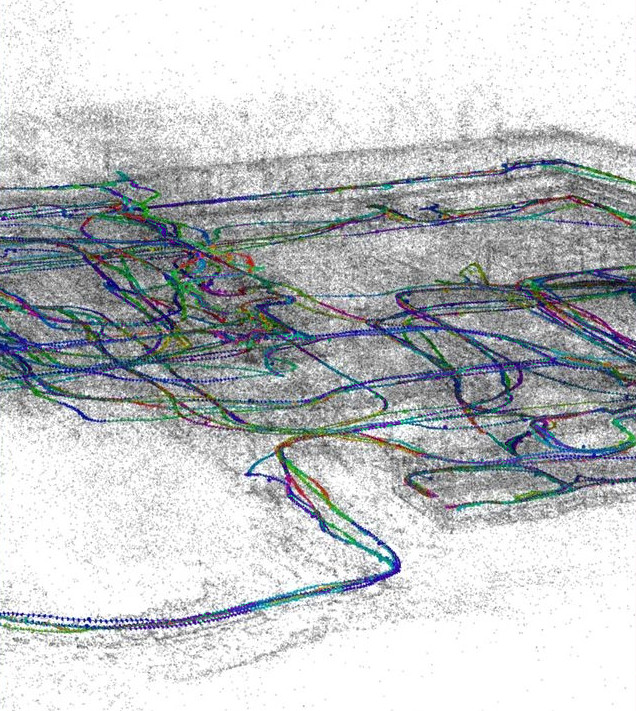}}\hfill%
\fbox{\includegraphics[width=0.325\linewidth]{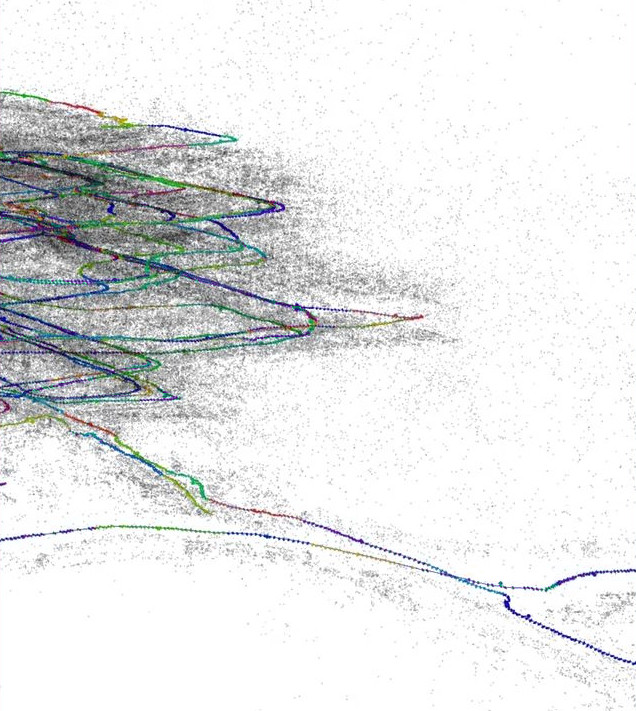}}%
}
\includegraphics[width=0.16\linewidth]{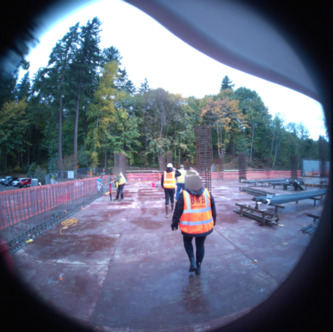}\hspace{0.0065\linewidth}%
\includegraphics[width=0.16\linewidth]{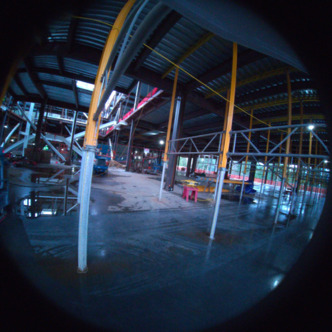}\hspace{0.01\linewidth}%
\includegraphics[width=0.16\linewidth]{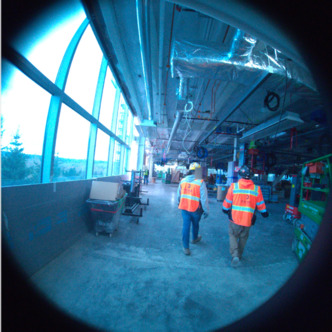}\hspace{0.0065\linewidth}%
\includegraphics[width=0.16\linewidth]{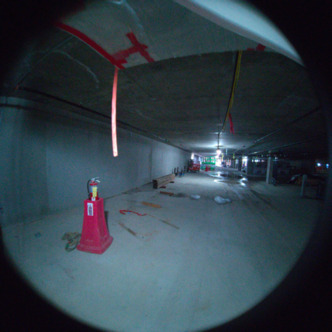}\hspace{0.01\linewidth}%
\includegraphics[width=0.16\linewidth]{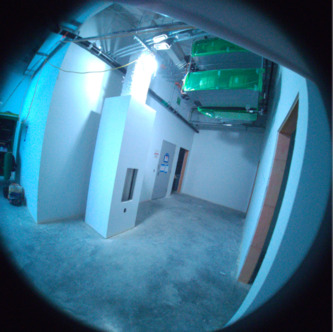}\hspace{0.0065\linewidth}%
\includegraphics[width=0.16\linewidth]{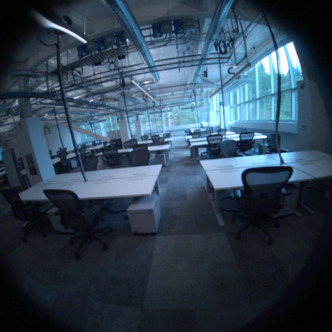}
    \caption{Top: A 3D map created from 275 Aria recordings (175 hours of data), captured over 15 months in a construction site, showing -- from left to right -- the state of the map after one, six, and fifteen months. The images at the bottom are samples from the respective points in time, depicting the transformation from an empty lot to an office building.}
    \label{fig:greenwich}
\end{figure}

\subsection{Life-long Mapping and Re-localization}
Precise 6DoF localization through SLAM or SfM is a common first step across many applications, as well as a base requirement for AR/VR world-locked rendering. It is a comparatively mature field, and Project Aria provides metric 6DoF trajectories as part of Aria MPS (see Sec.~\ref{sec:trajectories}). 
However, many challenges remain: One of these is to reliably re-localize across strong environment changes that occur in natural environments (typically due to lighting, weather, or human activity), as well as updating maps with such changes over time. 
Figure \ref{fig:greenwich} shows a map created from 275 Aria recordings, captured over 15 months, in a building that's being built: Using Aria as convenient recording device and building upon the per-recording trajectories from Aria MPS allows to focus on the core problem of long-range re-localization and map-updating under such strong environmental changes.

\begin{figure}
    \centering
{%
\setlength{\fboxsep}{0pt}%
\setlength{\fboxrule}{0.5pt}%
\includegraphics[width=0.495\linewidth]{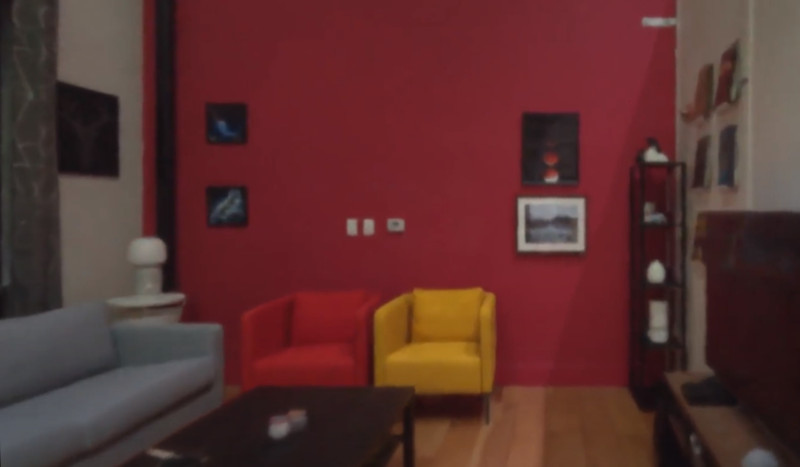}%
\hfill%
\includegraphics[width=0.495\linewidth]{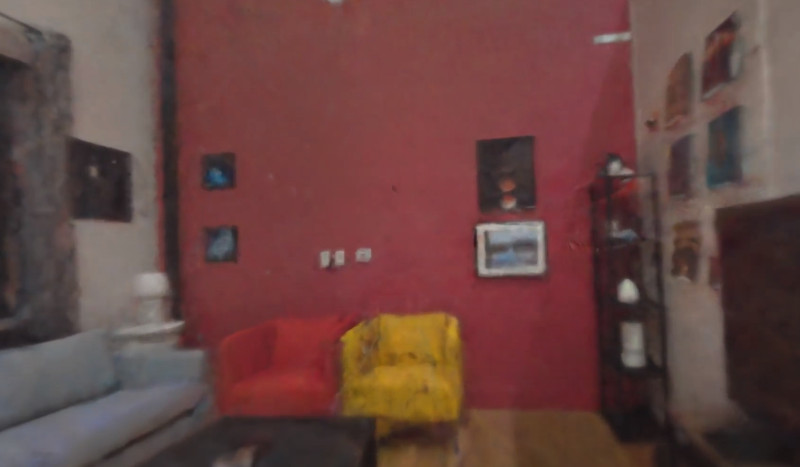}\\%
\fbox{\includegraphics[width=0.49\linewidth]{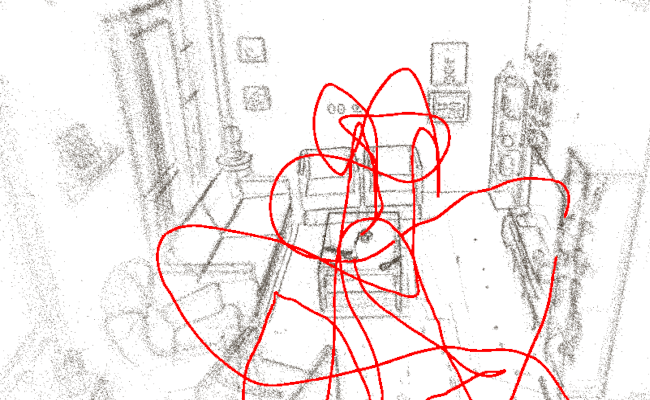}}%
\hfill%
\fbox{\includegraphics[width=0.49\linewidth]{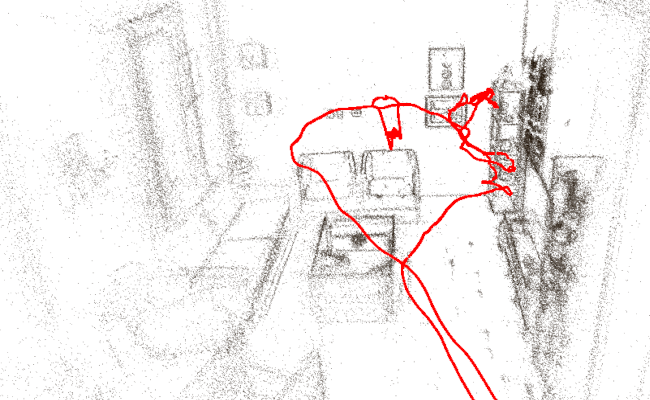}}
}%
    \caption{
   Two NeRF reconstructions obtained from Aria recordings using NerfStudio \cite{nerfstudio}. The left shows the result from a carefully curated, hand-held recording that covers the space well and avoids rapid motion. The right shows the result from an egocentric recording during a natural activity. The bottom figures visualize the raw pointcloud and trajectory from MPS. The resulting difference in quality is clearly visible.}
    \label{fig:nerfboxes}
\end{figure}

\subsection{Egocentric Scene Reconstruction and Understanding}
Reconstructing the surrounding scene and semantically identifying objects in it is a key problem for AR/VR applications -- from creating photo-realistic, virtual memories to identifying affordances of objects in the surroundings as part of a context-aware AI assistant. This becomes particularly challenging when the input data is not neatly curated or intentionally taken for this purpose, but rather stems from a form-factor and power-constrained wearable device undergoing natural, unconstrained human motion. 

Figure \ref{fig:nerfboxes} shows the results obtained by state-of-the-art methods for NeRF reconstruction \cite{nerfstudio} on Aria data, comparing careful ``scanning" motion with a natural activity.

\subsection{Object Interaction and Manipulation}
\begin{figure}
    \centering
\includegraphics[width=0.49\linewidth]{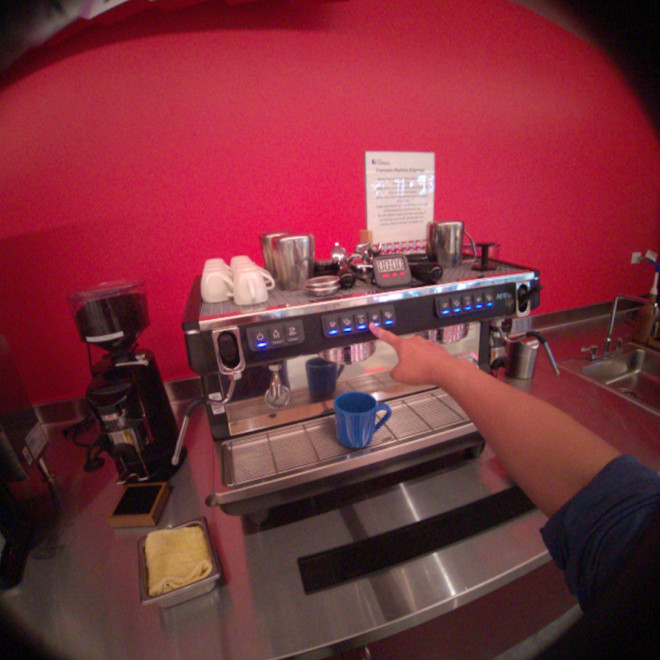}%
\hfill%
{%
\setlength{\fboxsep}{0pt}%
\setlength{\fboxrule}{0.5pt}%
\fbox{\includegraphics[width=0.49\linewidth]{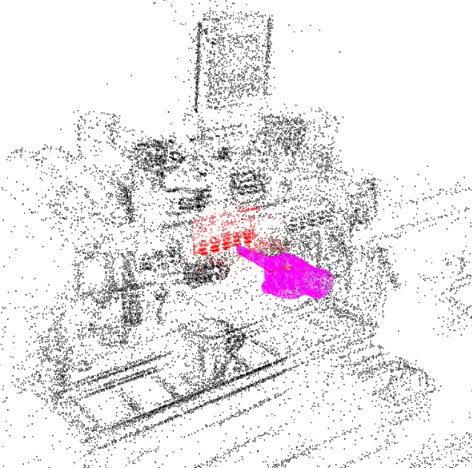}}}
    \caption{We use UmeTrack \cite{han2022umetrack} to track the articulated 3D hand-pose of the wearer in an Aria recording. Combined with the pointcloud from MPS, this allows to identify when the wearer touches a static object. The figures visualize this by coloring points within 8\,cm of the hand in red.}
    \label{fig:objectinteraction}
\end{figure}

Recognizing or tracking objects the user is interacting with, or identifying how the user is interacting with them, is another core egocentric machine perception task. It combines hand-tracking with object tracking, recognition, or scene understanding to connect \textit{things} with user intent or actions. 
Figure \ref{fig:objectinteraction} shows an example application that identifies when the user's hand is near an object in the scene. The approach uses all 3 cameras as well as the trajectory and pointcloud provided by MPS -- which can help in particular to resolve the otherwise common scale/depth ambiguity. Figure \ref{fig:mps_gaze} showed a similar approach using eye gaze to identify what the wearer is looking at.

\subsection{Activity Recognition and Attention}
\begin{figure}
    \centering
    \includegraphics[height=0.37\linewidth]{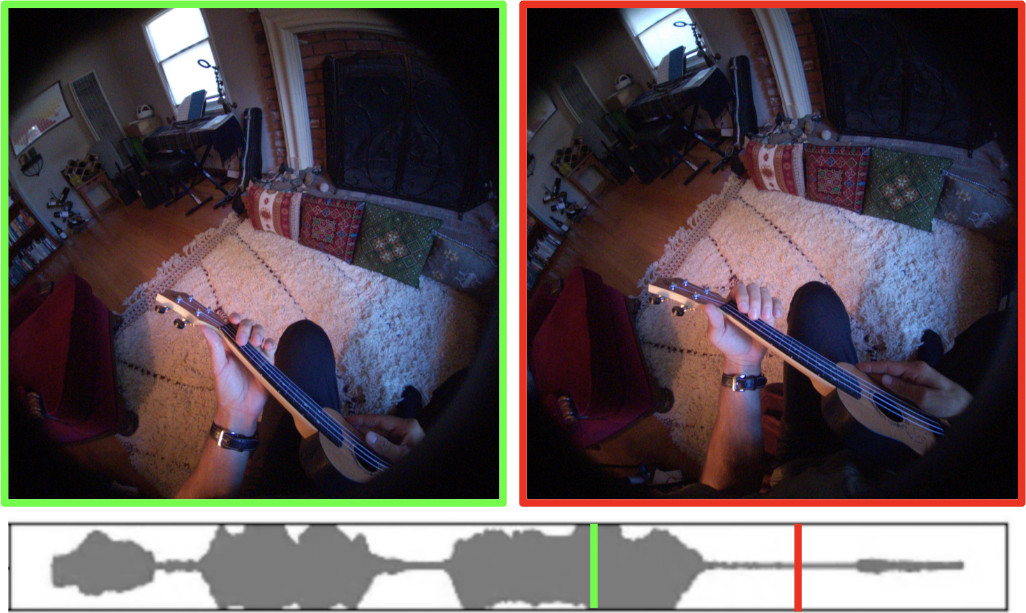}\hfill\includegraphics[height=0.37\linewidth]{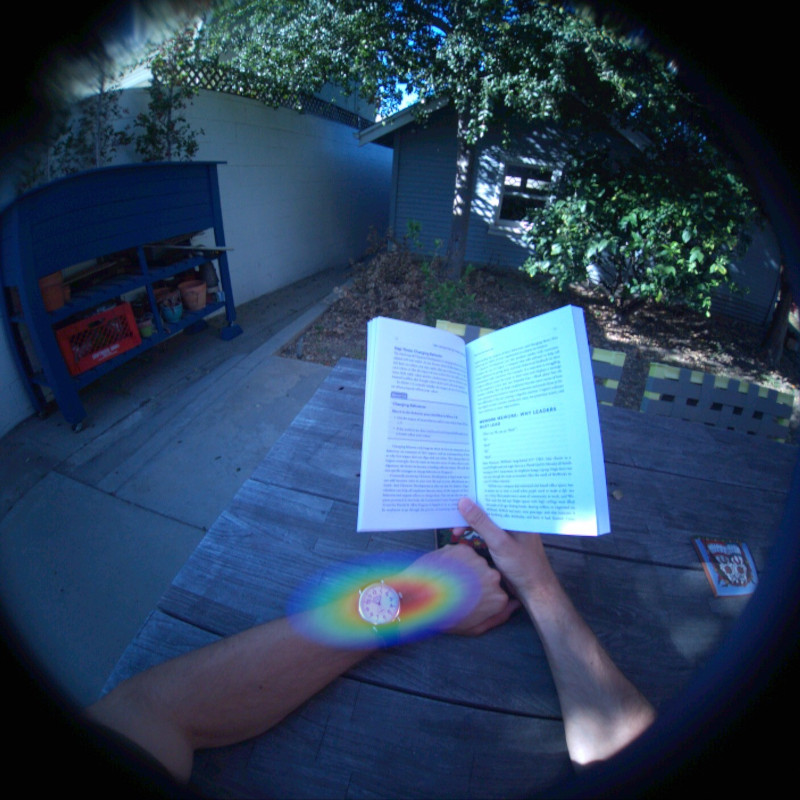}
    \caption{Left: two egocentric views of the wearer interacting with a guitar. The audio stream visualized below allows to disambiguate whether the wearer is actively playing or just holding the guitar. Right: the eye gaze (visualized as a heatmap) allows to distinguish whether the wearer is looking at the time or reading a book.}
    \label{fig:activity}
\end{figure}

Identifying what the wearer is doing or paying attention to is another likely component of any contextual AI assistant. While much information can be derived from egocentric images or videos alone, significant additional signal can be derived from other modalities, including spatial audio, motion, or eye gaze. 
Figure \ref{fig:activity} shows two example situations where these additional signals allow to disambiguate what the wearer is doing in otherwise ambiguous egocentric views.

\subsection{Summarization and Question Answering}
Summarization and Question Answering goes a step further than activity recognition, aiming to summarize relevant events and activities that occur over longer time periods and allowing to answer questions about them. Note that ``relevance" in this context is highly subjective and personal, making signals such as eye gaze or spatial audio key to select the most relevant information for the user. 

Furthermore, ``longer time periods" can vary from a few minutes to hours, days, and years -- with longer time-spans becoming increasingly important towards personalized AI assistants, but requiring datasets that do not currently exist. We believe that Project Aria's unique combination of form-factor and machine perception capabilities enables more research in this field towards egocentric, longitudinal summarization and question answering. 

%

\section{Conclusion}
\label{sec:conclusion}
With \ProjectAria{}, we introduce a new tool for the research community that can capture ecologically valid data as we expect it to be captured by future egocentric devices. We also make available a set of spatial AI machine perception technologies as a foundational building block for higher-level contextualized AI applications.
The data, and derived machine perception results, can provide the foundation for building novel compute and interaction paradigms needed in order to make AR successful and the rich integrated sensor suite provides unique opportunities to explore novel research, applications and use cases in a wide range of areas towards always-on contextualized AI.

\section*{Acknowledgements}

\ProjectAria{} was made possible by the contributions of the \ProjectAria{} team from Meta Reality Labs Research. We are indebted to the complete team and all partners of \ProjectAria{} who enabled its inception and continue to develop the platform.


\begin{thebibliography}{10}

\bibitem{ariadocs}
Project {A}ria {D}ocumentation.
\newblock \url{https://facebookresearch.github.io/projectaria_tools/}.

\bibitem{ariapilot}
Project {A}ria {P}ilot {D}ataset.
\newblock
  \url{https://facebookresearch.github.io/projectaria_tools/docs/open_datasets/pilot_dataset}.

\bibitem{ariatools}
Project {A}ria {T}ools on {G}it{H}ub.
\newblock \url{https://github.com/facebookresearch/projectaria_tools}.

\bibitem{ariawebsite}
Project {A}ria {W}ebsite.
\newblock \url{https://www.projectaria.com/}.

\bibitem{communityguidelines}
Project {A}ria {C}ommunity {G}uidelines.
\newblock
  \url{https://about.meta.com/realitylabs/projectaria/community-guidelines/}.

\bibitem{engel2014lsd}
Jakob Engel, Thomas Sch{\"o}ps, and Daniel Cremers.
\newblock {LSD-SLAM}: Large-scale direct monocular {SLAM}.
\newblock In {\em Computer Vision--ECCV 2014: 13th European Conference, Zurich,
  Switzerland, September 6-12, 2014, Proceedings, Part II 13}, pages 834--849.
  Springer, 2014.

\bibitem{touvron2023llama}
Hugo~Touvron et.al.
\newblock Llama 2: Open foundation and fine-tuned chat models, 2023.

\bibitem{han2022umetrack}
Shangchen Han, Po{-}Chen Wu, Yubo Zhang, Beibei Liu, Linguang Zhang, Zheng
  Wang, Weiguang Si, Peizhao Zhang, Yujun Cai, Tomas Hodan, Randi Cabezas, Luan
  Tran, Muzaffer Akbay, Tsz{-}Ho Yu, Cem Keskin, and Robert Wang.
\newblock Umetrack: Unified multi-view end-to-end hand tracking for {VR}.
\newblock In {\em {SIGGRAPH} Asia 2022 Conference Papers, {SA} 2022, Daegu,
  Republic of Korea, December 6-9, 2022}, 2022.

\bibitem{conf/icra/HarrisonN11}
Alastair Harrison and Paul Newman.
\newblock {TICSync}: Knowing when things happened.
\newblock In {\em 2011 IEEE International Conference on Robotics and
  Automation}, pages 356--363, 2011.

\bibitem{he2017mask}
Kaiming He, Georgia Gkioxari, Piotr Doll{\'a}r, and Ross Girshick.
\newblock Mask {R-CNN}.
\newblock In {\em Proceedings of the IEEE International Conference on Computer
  Vision}, pages 2961--2969, 2017.

\bibitem{kirillov2023segment}
Alexander Kirillov, Eric Mintun, Nikhila Ravi, Hanzi Mao, Chloe Rolland, Laura
  Gustafson, Tete Xiao, Spencer Whitehead, Alexander~C Berg, Wan-Yen Lo, et~al.
\newblock Segment anything.
\newblock {\em arXiv preprint arXiv:2304.02643}, 2023.

\bibitem{metari}
Meta {R}esponsible {I}nnovation {P}rinciples.
\newblock \url{https://about.meta.com/metaverse/responsible-innovation/}.

\bibitem{mourikis2007multi}
Anastasios~I Mourikis and Stergios~I Roumeliotis.
\newblock A multi-state constraint {K}alman filter for vision-aided inertial
  navigation.
\newblock In {\em Proceedings 2007 IEEE International Conference on Robotics
  and Automation}, pages 3565--3572. IEEE, 2007.

\bibitem{mur2017orb}
Raul Mur-Artal and Juan~D Tard{\'o}s.
\newblock {ORB-SLAM2}: An open-source slam system for monocular, stereo, and
  {RBG-D} cameras.
\newblock {\em IEEE Transactions on Robotics}, 33(5):1255--1262, 2017.

\bibitem{openai2023gpt4}
OpenAI.
\newblock {GPT-4} technical report.
\newblock {\em arXiv preprint arXiv:2303.08774}, 2023.

\bibitem{oquab2023dinov2}
Maxime Oquab, Timoth{\'e}e Darcet, Th{\'e}o Moutakanni, Huy Vo, Marc
  Szafraniec, Vasil Khalidov, Pierre Fernandez, Daniel Haziza, Francisco Massa,
  Alaaeldin El-Nouby, et~al.
\newblock {Dinov2}: Learning robust visual features without supervision.
\newblock {\em arXiv preprint arXiv:2304.07193}, 2023.

\bibitem{poole2022dreamfusion}
Ben Poole, Ajay Jain, Jonathan~T. Barron, and Ben Mildenhall.
\newblock {DreamFusion}: {Text-to-3D} using {2D} diffusion.
\newblock {\em arXiv}, 2022.

\bibitem{radford2021learning}
Alec Radford, Jong~Wook Kim, Chris Hallacy, Aditya Ramesh, Gabriel Goh,
  Sandhini Agarwal, Girish Sastry, Amanda Askell, Pamela Mishkin, Jack Clark,
  et~al.
\newblock Learning transferable visual models from natural language
  supervision.
\newblock In {\em International Conference on Machine Learning}, pages
  8748--8763. PMLR, 2021.

\bibitem{ramesh2022hierarchical}
Aditya Ramesh, Prafulla Dhariwal, Alex Nichol, Casey Chu, and Mark Chen.
\newblock Hierarchical text-conditional image generation with clip latents.
\newblock {\em arXiv preprint arXiv:2204.06125}, 2022.

\bibitem{smptet}
Linear {T}imecode.
\newblock \url{https://en.wikipedia.org/wiki/Linear_timecode}.

\bibitem{nerfstudio}
Matthew Tancik, Ethan Weber, Evonne Ng, Ruilong Li, Brent Yi, Justin Kerr,
  Terrance Wang, Alexander Kristoffersen, Jake Austin, Kamyar Salahi, Abhik
  Ahuja, David McAllister, and Angjoo Kanazawa.
\newblock Nerfstudio: A modular framework for neural radiance field
  development.
\newblock In {\em ACM SIGGRAPH 2023 Conference Proceedings}, SIGGRAPH '23,
  2023.

\bibitem{vrsdocs}
{VRS} {D}ocumentation.
\newblock \url{https://facebookresearch.github.io/vrs/docs/Overview}.

\end{thebibliography}

\end{document}